\documentclass[aps,prb,twocolumn,longbibliography,floatfix,10pt,superscriptaddress]{revtex4-2}
\usepackage{graphicx,latexsym}
\usepackage{amsmath,amssymb,amsfonts}
\usepackage{bm}
\usepackage{braket}
\usepackage{mathrsfs}

\newcommand{\up}{\uparrow}
\newcommand{\down}{\downarrow}
\newcommand{\m}{\mathcal}

\begin{document}
\title{Emergence of Charge-Imbalanced BCS State and Suppression of Nonequilibrium FFLO State in Asymmetric NSN Junctions}

\author{Taira Kawamura} 
\affiliation{Department of Physics, College of Science and Technology, Nihon University, Tokyo 101-8308, Japan}

\author{Yoji Ohashi}
\affiliation{Department of Physics, Keio University, 3-14-1 Hiyoshi, Kohoku-ku, Yokohama 223-8522, Japan}
\date{\today}

\begin{abstract}
We theoretically study nonequilibrium superconductivity in voltage-biased normal metal--superconductor--normal metal (NSN) junctions, focusing on effects of lead-coupling asymmetry and impurity scattering. Using the Keldysh Green’s function technique, we extend the thermal-equilibrium mean-field BCS theory to the case where the system is out of equilibrium, to analyze superconducting properties in the nonequilibrium steady state. We find that, in close analogy with the thermal-equilibrium case, the inhomogeneous nonequilibrium Fulde–Ferrell–Larkin–Ovchinnikov (NFFLO) state induced by nonequilibrium electron distributions is highly sensitive to impurity scattering, whereas the uniform nonequilibrium BCS (NBCS) state remains robust against nonmagnetic impurities. Moreover, lead-coupling asymmetry is also found to suppress the NFFLO phase and to split the NBCS phase into two distinct regimes, characterized by the presence or absence of a chemical-potential imbalance between quasiparticles and the condensate. We identify a phase transition or a crossover between these two NBCS states, as well as parameter regimes exhibiting bistability.
Our results provide a unified microscopic understanding of nonequilibrium superconductivity in NSN junctions under experimentally relevant conditions and are expected to provide a theoretical framework applicable to a broad class of nonequilibrium superconducting hybrid structures.
\end{abstract}

\maketitle
\section{Introduction \label{sec.Intro}}

Nonequilibrium superconductivity has been actively studied since the early stages of superconductivity research~\cite{GrayBook, TinkhamBook, Kopnin2001}. One of the key elements characterizing nonequilibrium superconductivity is the electron distribution. In the thermal-equilibrium state, it is given by the Fermi–Dirac function, which is uniquely determined by the temperature and chemical potential. In the nonequilibrium superconducting state, by contrast, it no longer obeys the Fermi–Dirac form. The electron distribution thus constitutes an additional degree of freedom in the nonequilibrium case and plays crucial roles in determining superconducting properties. Nonequilibrium phenomena, such as charge imbalance~\cite{Clarke1972, Tinkham1972_PRL, Tinkham1972, Pethick1979, Pethick1980, Schmid1975} and the enhancement of superconductivity under microwave irradiation~\cite{Wyatt1966, Dayem1967, Ivlev1973, Kommers1977, Pals1979, Gorkov1969, Eliashberg1970, Tikhonov2018, Tikhonov2020_review}, can only be understood by properly accounting for the nonequilibrium electron distribution.

\begin{figure}[t]
\centering
\includegraphics[width=\columnwidth]{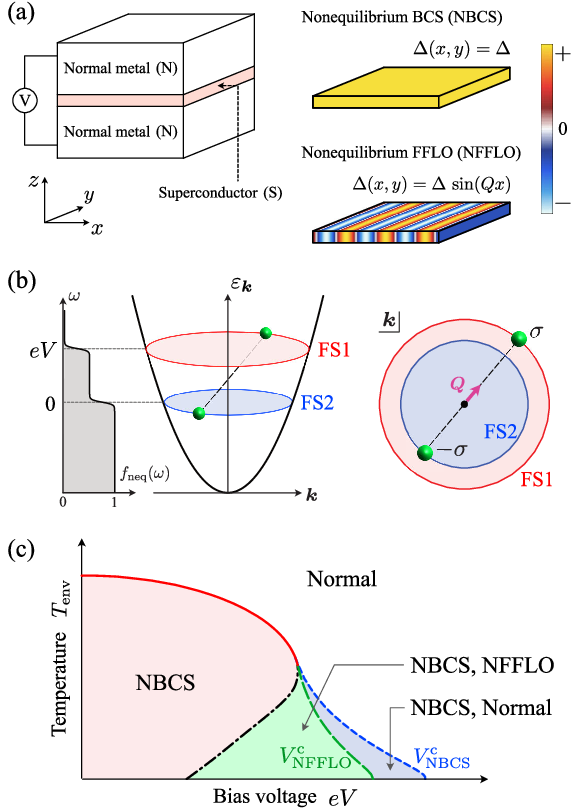}
\caption{(a) Model NSN junction~\cite{Kawamura2024}. A thin superconducting layer is sandwiched between two normal-metal leads and is driven out of equilibrium by a bias voltage $V$ applied between the leads. The superconducting order parameter is uniform in the nonequilibrium BCS (NBCS) state, whereas it exhibits a spontaneous spatial modulation in the $x$–$y$ plane in the nonequilibrium FFLO (NFFLO) state. (b) Schematic illustration of the FFLO-type Cooper-pairing mechanism in the nonequilibrium superconductor. The two-step electron distribution in Eq.~\eqref{eq.fneq.twostep} gives rise to two effective Fermi surfaces (FS1 and FS2), leading to Cooper pairs with a nonzero center-of-mass momentum $\bm{Q}$. (c) Schematic phase diagram of a nonequilibrium superconductor in the clean limit with symmetric superconductor–lead couplings~\cite{Kawamura2024}. The system exhibits bistability in the green and blue shaded regions.}
\label{Fig.NSN}
\end{figure}

In our recent paper~\cite{Kawamura2024}, we clarified that nonequilibrium electron distributions can induce spatially nonuniform superconducting states. We considered a normal metal--superconductor--normal metal (NSN) junction, illustrated in Fig.~\ref{Fig.NSN}(a), to study nonequilibrium superconductivity driven by a bias voltage $V$ applied between the two normal-metal leads. When the superconductor is sufficiently thin along the $z$ direction, so that the relaxation time due to electron tunneling into the leads is much shorter than those associated with inelastic electron–phonon and electron–electron scattering, the nonequilibrium electron distribution in the superconductor is dominated by the coupling to the leads~\cite{Kawamura2024, Kawamura2025, Tserkovnyak2002}. In particular, when the two couplings between the superconductor (S) and the leads (N) are symmetric, it takes the two-step form~\cite{Kawamura2024},
\begin{equation}
f_{\rm neq}(\omega) \simeq \frac{1}{2}\bigl[f(\omega - eV) + f(\omega)\bigr],
\label{eq.fneq.twostep}
\end{equation}
where $f(\omega) = [1 + e^{\omega/T_{\rm env}}]^{-1}$ is the Fermi–Dirac distribution function in the thermal-equilibrium normal-metal leads at temperature $T_{\rm env}$. We briefly note that such two-step nonequilibrium electron distributions have been observed in various mesoscopic systems under an applied bias voltage~\cite{Pothier1997, Anthore2003, Huard2005, Franceschi2002, Tikhonov2020, Chen2009, Bronn2013}.

In general, multistep structures in nonequilibrium electron distributions work like multiple effective ``Fermi surfaces"~\cite{Kawamura2020, Kawamura2022, Kawamura2023, Kawamura2024_AAPPS, Abanin2005, Ono2019, Shi2024}. The two-step distribution in Eq.~\eqref{eq.fneq.twostep} gives rise to two effective Fermi surfaces of different sizes in momentum space, denoted as FS1 and FS2, as schematically shown in Fig.~\ref{Fig.NSN}(b). In this situation, Cooper pairing can occur between electrons around different effective Fermi surfaces, leading to Cooper pairs with a nonzero center-of-mass momentum $\bm{Q}$. As a result, an inhomogeneous superconducting state is stabilized, where the superconducting order parameter exhibits a spontaneous spatial modulation in the $x$–$y$ plane, as illustrated in Fig.~\ref{Fig.NSN}(a). This state is reminiscent of the Fulde–Ferrell–Larkin–Ovchinnikov (FFLO) state in thermal-equilibrium superconductors under an external magnetic field~\cite{Fulde1964, Larkin1964, Matsuda2007}, and is thus referred to as the nonequilibrium FFLO (NFFLO) state. Importantly, the NFFLO state can be stabilized even in the absence of an external magnetic field.

In Ref.~\cite{Kawamura2024}, we showed that the NFFLO state can be realized as a stable nonequilibrium steady state in a voltage-biased NSN junction. In the schematic phase diagram in Fig.~\ref{Fig.NSN}(c), the red region represents the nonequilibrium uniform BCS-type superconducting phase (NBCS), while the NFFLO state is stabilized in the green region. Here, we briefly note that the system actually exhibits bistability in the green region (and also in the blue region): When the bias voltage is increased from $V=0$, the NBCS state remains stable up to the critical voltage $V_{\rm NBCS}^{\rm c}$. On the other hand, when the voltage is decreased from the normal phase (white region), the system undergoes a transition into the NFFLO state at $V_{\rm NFFLO}^{\rm c}$.

We point out that the above-mentioned results were obtained in the case where the superconductor–lead couplings are symmetric. In realistic devices, however, such an ideal situation is not always realized, and some degree of coupling asymmetry is usually unavoidable. Since such asymmetry directly modifies the nonequilibrium electron distribution from the symmetric case in Eq.~\eqref{eq.fneq.twostep}, it can affect nonequilibrium superconducting properties. Moreover, in real systems, the relaxation time due to electron–impurity scattering can be comparable to the electron escape time from the superconductor into the normal-metal leads. In this case, we also need to include impurity-scattering effects on the stability of nonequilibrium superconducting states. Indeed, the thermal-equilibrium FFLO state is known to be highly sensitive to impurity scattering~\cite{Matsuda2007}, implying that the NFFLO state is likewise susceptible to impurities. Nevertheless, a systematic understanding of how these experimentally unavoidable factors modify the nonequilibrium superconducting phase diagram remains lacking.

In this paper, we elucidate how impurity scattering and asymmetry of the superconductor–lead couplings affect nonequilibrium superconductivity in an NSN junction. For this purpose, we extend the Thouless criterion~\cite{Thouless1960, Kadanoff1961, KadanoffBaym1962} and the mean-field BCS theory~\cite{BCS1957}, originally developed for thermal-equilibrium superconductors, to the nonequilibrium setting using the nonequilibrium Green’s function technique. Our main findings are summarized as follows. (1) Both impurity scattering and coupling asymmetry suppress the NFFLO state. (2) The NBCS state is robust against nonmagnetic impurities but fragile with respect to magnetic impurities, in close analogy with the thermal-equilibrium BCS state~\cite{Anderson1959}. (3) In the presence of asymmetric superconductor–lead couplings, the NBCS state can be classified into two types, distinguished by the presence or absence of a chemical-potential imbalance between quasiparticles and the condensate. A phase transition or crossover can occur between these two NBCS states.

This paper is organized as follows. In Sec.~\ref{sec.Model}, we introduce our model Hamiltonian. In Sec.~\ref{sec.formalism}, we extend the Thouless criterion and the mean-field BCS theory to the nonequilibrium steady state. Using these theoretical frameworks, we examine in Sec.~\ref{sec.FFLO.asy} how asymmetry of the superconductor–lead couplings affects the NFFLO state. In Sec.~\ref{sec.NBCS}, we discuss the properties of the NBCS state under asymmetric lead couplings. Impurity effects on the NFFLO and NBCS states are discussed in Sec.~\ref{sec.imp*}. Throughout this paper, we set $\hbar = k_{\rm B} = 1$, and the volumes of the two normal-metal leads are taken to be unity for simplicity.

\section{Model Hamiltonian \label{sec.Model}}

We consider a thin superconducting layer sandwiched between two normal-metal leads, as schematically shown in Fig.~\ref{Fig.NSN}(a). This device is modeled by the Hamiltonian~\cite{Kawamura2024}
\begin{equation}
H = H_{\rm sys} + H_{\rm lead} + H_{\rm mix}.
\label{eq.model.Hamiltonian}
\end{equation}
Here, the Hamiltonian of the nonequilibrium superconductor (S) is given by
\begin{align}
H_{\rm sys} &= 
\sum_{\bm{k},\sigma}
\big[\xi_{\bm{k}} +e\varphi\big] a^\dagger_{\bm{k},\sigma} a_{\bm{k},\sigma} 
\notag\\
&\hspace{0.5cm}
- U\sum_{\bm{k},\bm{k},\bm{q}} a^\dagger_{\bm{k}+\bm{q}/2,\up} a^\dagger_{-\bm{k}+\bm{q}/2, \down} a_{-\bm{k}'+\bm{q}/2, \down} a_{\bm{k}'+\bm{q}/2,\up}
\notag\\
&\hspace{0.5cm}
+
H_{\rm imp}
+
H_{\rm mag}.
\label{eq.Hsys.def}	
\end{align}
In Eq.~\eqref{eq.Hsys.def}, the operator $a^\dagger_{\bm{k},\sigma}$ creates an electron with momentum $\bm{k}$ and spin $\sigma=\up,\down$. $\xi_{\bm{k}}=\bm{k}^2/(2m) -\mu_{\rm sys}$ is the kinetic energy and $\varphi$ is the electrostatic potential of the superconducting region. The attractive interaction $-U$ ($<0$) gives rise to the Cooper pairing. The last two terms, $H_{\rm imp}$ and $H_{\rm mag}$, describe elastic scattering from nonmagnetic and magnetic impurities, respectively. They are given by
\begin{align}
&
H_{\rm imp} =
u_{\rm imp} \sum_{\bm{k},\bm{q}} \sum_{j=1}^{N_{\rm imp}}\,
e^{-i\bm{q}\cdot \bm{R}^{\rm i}_j}\,
a^\dagger_{\bm{k}+\bm{q},\sigma} a_{\bm{k},\sigma}
\label{eq.H.imp}
,\\
&
H_{\rm mag} = 
u_{\rm mag}\sum_{\sigma,\sigma'}\sum_{\bm{k},\bm{q}} 
\sum_{j=1}^{N_{\rm mag}}\,
e^{-i\bm{q}\cdot \bm{R}^{\rm m}_j}\,
a^\dagger_{\bm{k}+\bm{q},\sigma} 
\bm{\sigma}_{\sigma,\sigma'}
a_{\bm{k},\sigma'}
\cdot
\bm{S}_j.
\end{align}
Here, $u_{\rm imp}$ ($u_{\rm mag}$) denotes the scattering strength of nonmagnetic (magnetic) impurities, and $N_{\rm imp}$ ($N_{\rm mag}$) is the corresponding number of impurities. The vector $\bm{R}_j^{\rm i}$ ($\bm{R}_j^{\rm m}$) is the spatial position of the $j$th impurity, which is assumed to be randomly disributed. The vector $\bm{\sigma}=(\sigma_x,\sigma_y,\sigma_z)$ are the Pauli matrices acting on the electron spin degrees of freedom. $\bm{S}_j$ represents the classical spin of the $j$th magnetic impurity.

\begin{figure}[t]
\centering
\includegraphics[width=\columnwidth]{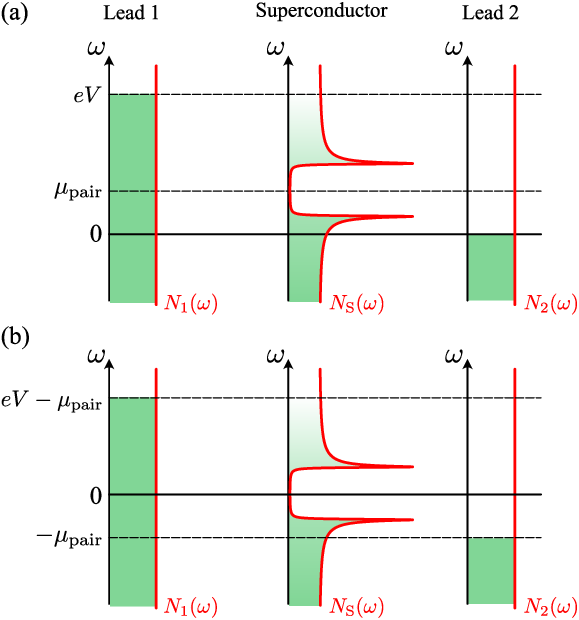}
\caption{Schematic energy diagram of our model in the nonequilibrium steady state. (a) The energy reference is taken at the Fermi level of the lead 2. The lead 1 is filled up to energy $eV$, and the bias voltage $V$ drives the superconductor out of equilibrium. (b) The energy is measured from the pair chemical potential $\mu_{\rm pair}$, which is defined in Eq.~\eqref{eq.OP}. This choice of energy reference eliminates the trivial time dependence of the superconducting order parameter (see Sec.~\ref{sec.formalism.NBCS} for details). Unless otherwise stated, results are presented using the energy reference in panel (a), whereas the reference in panel (b) is used in formulating the nonequilibrium BCS theory.}
\label{Fig.energy}
\end{figure}

The normal-metal leads (N) in Fig.~\ref{Fig.NSN}(a) are described by the Hamiltonian~\cite{Jauho1994, Kawamura2024}
\begin{equation}
H_{\rm lead}
=
\sum_{\alpha=1,2}
\sum_{\bm{p},\sigma}
\xi^\alpha_{\bm{p}}(t)\,
c^{\alpha \dagger}_{\bm{p},\sigma}
c^\alpha_{\bm{p},\sigma},
\end{equation}
where $c^{\alpha \dagger}_{\bm{p},\sigma}$ creates an electron with momentum $\bm{p}$ and spin $\sigma$ in the lead $\alpha$($=1,2$), with $\xi^\alpha_{\bm{p}}(t)=\bm{p}^2/(2m)-\mu_\alpha(t)$ being the kinetic energy. Each lead is assumed to be maintained in the thermal equilibrium state at the temperature $T_{\rm env}$ and the electrochemical potential $\mu_\alpha(t)$. The time dependence of $\mu_\alpha(t)$ models the applied bias voltage that drives the superconducting region (S) in Fig. 1(a) out of equilibrium, while the leads themselves are assumed to relax sufficiently fast to remain in equilibrium at all times. The difference between the electrochemical potentials of the two leads, $\mu_2(t)-\mu_1(t)$, corresponds to the bias voltage $eV$ applied between the leads, as schematically illustrated in Fig.~\ref{Fig.energy}. To model the quench into the nonequilibrium state, we assume the following time dependence of the lead electrochemical potentials:
\begin{subequations}
\begin{align}
\mu_1(t) &= \mu_{\rm lead} - eV\,\Theta(t), \\
\mu_2(t) &= \mu_{\rm lead}.
\end{align}
\end{subequations}
Here, $\Theta(t)$ is the step function. At $t=0$, the entire system is in thermal equilibrium with $\mu_1=\mu_2=\mu_{\rm lead}$. For $t>0$, a non-zero bias voltage $V$ is applied between the two normal-metal leads, and the superconducting region is driven out of equilibrium. In the following, we focus on the nonequilibrium steady state reached at long times ($t\to \infty$), after the bias voltage $V$ is switched on.

The electron tunneling between the superconductor (S) and the normal-metal leads (N) is described by 
\begin{equation}
H_{\rm mix} = \sum_{\alpha=1,2} \sum_{\bm{p},\bm{k},\sigma}
\big[\m{T}_\alpha c^{\alpha \dagger}_{\bm{p},\sigma} a_{\bm{k},\sigma} +{\rm H.c.} \big],
\label{eq.H.lead}
\end{equation}
where $\m{T}_\alpha$ is the tunneling matrix element between the superconductor and lead $\alpha$. Experimentally, the magnitude of $\m{T}_\alpha$ is determined by microscopic details of the junction interface, such as the contact quality and the strength of the insulating barrier. While Ref.~\cite{Kawamura2024} only considers the simplest symmetric case ($\m{T}_1=\m{T}_2$), we also deal with the asymmetric case ($\m{T}_1\neq \m{T}_2$) in this paper.

We briefly comment on the connection between our model in Eq.~\eqref{eq.model.Hamiltonian} and experimentally accessible systems. The NSN sandwich structures have been used in early experiments on the so-called charge imbalance phenomenon in nonequilibrium superconductors~\cite{Clarke1972}, where Cu (N) and Al (S) strips are stacked. More recently, in the field of spintronics, sandwich-type ferromagnet–superconductor–ferromagnet (FSF) junctions,  resembling Fig.~\ref{Fig.NSN}(a), have also been realized by using CoFe (F) and Al (S)~\cite{Yang2010}. In these devices, the thickness of Al layer is approximately 4.5~nm, which is sufficiently small compared to the typical electron–phonon and electron–electron scattering mean free path in Al~\cite{Tserkovnyak2002}. Replacing the ferromagnetic electrodes by normal metals such as Cu would thus realize the NSN junction considered in this work.

\section{Formalism \label{sec.formalism}}

In this section, we explain nonequilibrium extension of the Thouless criterion in Sec.~\ref{sec.Thouless}, as well as that of the BCS mean-field theory in Sec.~\ref{sec.formalism.NBCS}, which together provide a complementary framework for analyzing the superconducting steady state of the NSN junction in the presence of asymmetric superconductor-lead couplings and impurity scattering.

The Thouless criterion characterizes the onset of superconductivity from the normal state by examining the divergence of the superconducting pair susceptibility~\cite{Thouless1960, Kadanoff1961, KadanoffBaym1962}. Within this framework, we study the stability of the NFFLO state against the lead-coupling asymmetry, as well as impurity scattering.

By contrast, the nonequilibrium BCS mean-field theory allows us to directly study the symmetry-broken phase. Using this framework, we investigate the nonequilibrium properties of the NBCS state.

\subsection{Nonequilibrium Thouless criterion \label{sec.Thouless}}

We introduce the $2\times2$ Keldysh Green's function for electrons with spin $\sigma$ in the superconductor,
\begin{equation}
\hat{G}_\sigma(\bm{k},\omega)
=
\begin{pmatrix}
G^{\rm R}_\sigma(\bm{k},\omega) &
G^{\rm K}_\sigma(\bm{k},\omega) \\[4pt]
0 &
G^{\rm A}_\sigma(\bm{k},\omega)
\end{pmatrix},
\label{eq.Ghat}
\end{equation}
where the superscripts R, A, and K denote the retarded, advanced, and Keldysh components, respectively. Each matrix element in Eq.~\eqref{eq.Ghat} is defined as
\begin{subequations}
\begin{align}
G^{\rm R}_\sigma(\bm{k},\omega)
&=
-i \int_{-\infty}^\infty dt\, e^{i\omega t}\,
\Theta(t)
\braket{
\big[
a_{\bm{k},\sigma}(t),
a^\dagger_{\bm{k},\sigma}(0)
\big]_+
},
\\
G^{\rm A}_\sigma(\bm{k},\omega)
&=
\big[G^{\rm R}_\sigma(\bm{k},\omega)\big]^*,
\\[4pt]
G^{\rm K}_\sigma(\bm{k},\omega)
&=
-i \int_{-\infty}^\infty dt\, e^{i\omega t}
\braket{
\big[
a_{\bm{k},\sigma}(t),
a^\dagger_{\bm{k},\sigma}(0)
\big]_-
},
\end{align}
\end{subequations}
with $[A,B]_\pm = AB \pm BA$. In the nonequilibrium steady state, the Green's function satisfies the following Dyson equation~\cite{Stefanucci2013, Rammer2007, Haug2008}:
\begin{equation}
\hat{G}_\sigma(\bm{k},\omega)
=
\hat{G}_{0,\sigma}(\bm{k},\omega)
+
\hat{G}_{0,\sigma}(\bm{k},\omega)
\hat{\Sigma}_\sigma(\bm{k},\omega)
\hat{G}_\sigma(\bm{k},\omega),
\label{eq.Dyson}
\end{equation}
where
\begin{align}
&
\hat{G}_{0,\sigma}(\bm{k},\omega)
\notag\\
&=
\scalebox{0.95}{$\displaystyle
\begin{pmatrix}
\dfrac{1}{\omega +i\delta - \xi_{\bm{k}} -e\varphi} &
-2\pi i\,\delta(\omega-\xi_{\bm{k}}-e\varphi)
\big[1-2f_{\rm ini}(\omega)\big] \\
0 &
\dfrac{1}{\omega -i\delta - \xi_{\bm{k}} -e\varphi}
\end{pmatrix} $}
\label{eq.G0}
\end{align}
is the Green's function of a non-interacting electron gas in the absence of the coupling with the leads, as well as impurity scattering. Here, $\delta>0$ is an infinitesimally small positive number. Importantly, the final steady-state superconducting Green's function does not depend on the initial electron distribution $f_{\rm ini}(\omega)$, as will be shown later.

The self-energy $\hat{\Sigma}_\sigma$ in the Dyson equation~\eqref{eq.Dyson} consists of contributions from the normal-metal leads and from scattering by nonmagnetic impurities,
\begin{equation}
\hat{\Sigma}_{\sigma}(\bm{k},\omega)
=
\hat{\Sigma}_{{\rm lead},\sigma}(\bm{k},\omega)
+
\hat{\Sigma}_{{\rm imp},\sigma}(\bm{k},\omega).
\label{eq.self}
\end{equation}
At this stage, we ignore magnetic impurity scattering, which will later be taken into account when we extend the BCS mean-field theory in Sec.~\ref{sec.formalism.NBCS}.

Within the second-order Born approximation with respect to the tunneling matrix element $\m{T}_\alpha$, the self-energy $\hat{\Sigma}_{{\rm lead},\sigma}$ in Eq.~\eqref{eq.self} is given by~\cite{Kawamura2020}
\begin{align} 
\hat{\Sigma}_{{\rm lead},\sigma}(\bm{k},\omega) &= 
\sum_{\alpha=1,2} 
\begin{pmatrix} 
- i\dfrac{\gamma_\alpha}{2} 
& -i\gamma_\alpha\big[1 -2f(\omega-eV\delta_{\alpha,1})\big] \\[6pt]
0 & i\dfrac{\gamma_\alpha}{2} 
\end{pmatrix} 
\notag\\[4pt]
&\equiv 
\begin{pmatrix} 
-i\dfrac{\gamma}{2} & 
-i\gamma \big[ 1 -2f_{\rm neq}(\omega) \big] \\ 
0 & i\dfrac{\gamma}{2} 
\end{pmatrix}. 
\label{eq.self.lead} 
\end{align}
Here,
\begin{equation} 
f(\omega) = \frac{1}{e^{\omega/T_{\rm env}} +1} 
\end{equation} 
is the Fermi-Dirac distribution function in the thermal-equilibrium leads, and \begin{subequations} 
\begin{align} 
& \gamma_{\alpha} = 2\pi N_\alpha(0) |\m{T}_\alpha|^2 \label{eq.gamma.lead} ,\\[4pt]
& \gamma = \gamma_1 +\gamma_2 ,\\[4pt] 
& f_{\rm neq}(\omega) = 
\frac{\gamma_1}{\gamma}f(\omega-eV) +\frac{\gamma_2}{\gamma} f(\omega). 
\label{eq.fneq.def}
\end{align} 
\end{subequations} 
In Eq.~\eqref{eq.gamma.lead}, $N_\alpha(\omega)$ is the single-particle density of states in the lead $\alpha$ $(=1,2)$. In deriving the self-energy $\hat{\Sigma}_{\rm lead}$, we have assumed that $N_\alpha(\omega)$ is not affected by the proximity effect and have ignored its energy dependence in the vicinity of $\omega=0$, as $N_\alpha(\omega)\simeq N_\alpha(0)$~\cite{Kawamura2024, Moor2009, Snyman2009}. In the following, we use $\gamma_\alpha$ as a parameter characterizing the coupling strength between the superconductor and the lead $\alpha$. The asymmetry of the lead couplings is then quantified by
\begin{equation}
P_{\rm lead}=
\frac{|\gamma_1-\gamma_2|}{\gamma_1+\gamma_2}.
\label{eq.Plead}
\end{equation}

We treat the impurity self-energy $\hat{\Sigma}_{{\rm imp},\sigma}$ in Eq.~\eqref{eq.self} within the self-consistent Born approximation~\cite{Rammer2007, Mahan2000}, which yields
\begin{equation}
\hat{\Sigma}_{{\rm imp},\sigma}(\bm{k},\omega)
=
\begin{pmatrix}
-\dfrac{i}{2\tau_{\rm imp}} &
\dfrac{1}{\tau_{\rm imp}}
\big[1-2f_{\rm neq}(\omega)\big] \\
0 &
\dfrac{i}{2\tau_{\rm imp}}
\end{pmatrix}.
\label{eq.self.imp}
\end{equation}
Here,
\begin{equation}
\frac{1}{\tau_{\rm imp}}=
2\pi N(0)\,n_{\rm imp}u_{\rm imp}^2
\label{eq.def.tau.imp}
\end{equation}
is the elastic impurity scattering rate, where $N(0)$ is the normal-state density of states of the superconductor, and $n_{\rm imp}=N_{\rm imp}/V$ is the impurity concentration.

Substituting the self-energy corrections in Eqs.~\eqref{eq.self.lead} and \eqref{eq.self.imp} into the Dyson equation~\eqref{eq.Dyson}, one obtains the single-particle Green's functions as
\begin{subequations}
\begin{align}
&
G^{\rm R(A)}_\sigma(\bm{k},\omega)=
\frac{1}{\omega \pm i\tilde{\gamma}/2 - \xi_{\bm{k}} -e\varphi},
\label{eq.GR.imp}
\\[4pt]
&
G^{\rm K}_\sigma(\bm{k},\omega)=
-\frac{i\tilde{\gamma}\big[1-2f_{\rm neq}(\omega)\big]}
{[\omega-\xi_{\bm{k}} -e\varphi]^2+\tilde{\gamma}^2/4},
\end{align}
\label{eq.G.imp}
\end{subequations}
where the effective linewidth is defined as
\begin{equation}
\tilde{\gamma}
=
\gamma + \frac{1}{\tau_{\rm imp}}.
\label{eq.til.gamma}
\end{equation}

According to the Thouless criterion~\cite{Thouless1960, Kadanoff1961, KadanoffBaym1962}, a Cooper instability is signaled by the appearance of a pole in the retarded particle–particle scattering $T$-matrix $\Gamma^{\rm R}(\bm{q},\nu)$ at $\bm{q}=\bm{Q}$ and $\nu=2\mu_{\rm pair}$. Here, $\bm{Q}$ is the Cooper-pair center-of-mass momentum and $\mu_{\rm pair}$ is the pair chemical potential. Within the mean-field (ladder) approximation, the retarded $T$-matrix is given by~\cite{Kawamura2020}
\begin{equation}
\Gamma^{\rm R}(\bm{q},\nu)
=
\frac{-U}{1+U\Pi^{\rm R}(\bm{q},\nu)}.
\label{eq.Tmat}
\end{equation}
Accordingly, the Thouless criterion is expressed as
\begin{align}
& 1 + U\, {\rm Re}\big[\Pi^{\rm R}(\bm{Q},2\mu_{\rm pair})\big] = 0,
\label{eq.Thouless.Re}
\\[4pt]
& {\rm Im}\big[\Pi^{\rm R}(\bm{Q},2\mu_{\rm pair})\big] = 0.
\label{eq.Thouless.Im}
\end{align}
Here, $\Pi^{\rm R}$ is the retarded pair (particle–particle) correlation function, having the form,
\begin{widetext}
\begin{align}
\Pi^{\rm R}(\bm{q},\nu)
&=
\frac{i}{2}\sum_{\bm{k}}\int_{-\infty}^\infty \frac{d\omega}{2\pi}
\Big[
G^{\rm R}_\up(\bm{k}+\bm{q}/2, \omega+\nu)
G^{\rm K}_\down(-\bm{k}+\bm{q}/2, -\omega)
\Lambda(\bm{q},\omega+\nu,-\omega)
\notag\\
&\hspace{5cm}+
G^{\rm K}_\up(\bm{k}+\bm{q}/2, \omega+\nu)
G^{\rm R}_\down(-\bm{k}+\bm{q}/2, -\omega)
\Lambda^*(\bm{q},\omega+\nu,-\omega)
\Big]
\notag\\
&=
\frac{N(0)}{4}
\int_{-\infty}^\infty d\omega
\Bigg[
\frac{1-2f_{\rm neq}(-\omega)}{2\bar{q}}
\log\!\left(
\frac{\omega+\nu/2+i\tilde{\gamma}/2+\bar{q}}
{\omega+\nu/2+i\tilde{\gamma}/2-\bar{q}}
\right)
\Lambda(\bm{q},\omega+\nu,-\omega)
\notag\\
&\hspace{5cm}
-
\frac{1-2f_{\rm neq}(\omega+\nu)}{2\bar{q}}
\log\!\left(
\frac{\omega+\nu/2-i\tilde{\gamma}/2+\bar{q}}
{\omega+\nu/2-i\tilde{\gamma}/2-\bar{q}}
\right)
\Lambda^*(\bm{q},\omega+\nu,-\omega)
\Bigg],
\label{eq.PiR}
\end{align}
where $\bar{q}=v_{\rm F}|\bm{q}|/2$, with $v_{\rm F}$ being the Fermi velocity. The function $\Lambda$ in Eq.~\eqref{eq.PiR} represents the vertex correction arising from impurity scattering. To ensure consistency with the Ward–Takahashi identity, the vertex function $\Lambda$ incorporates the impurity ladder diagrams~\cite{Rammer2007, Rammer2018}. Accordingly, $\Lambda$ satisfies
\begin{equation} 
\Lambda(\bm{q}, \omega+\nu, -\omega)
=
1
+
n_{\rm imp} u_{\rm imp}^2
\sum_{\bm{k}}
G^{\rm R}_\up(\bm{k}+\bm{q}/2,\omega +\nu)
G^{\rm A}_\down(-\bm{k}+\bm{q}/2, -\omega)\,
\Lambda(\bm{q}, \omega+\nu, -\omega).
\label{eq.BSeq} 
\end{equation}
The derivation of Eqs.~\eqref{eq.PiR} and \eqref{eq.BSeq} are summarized in Appendix~\ref{sec.app.Pi}. Substituting Eq.~\eqref{eq.GR.imp} into Eq.~\eqref{eq.BSeq} and carrying our the $\bm{k}$ summation, one has
\begin{equation}
\Lambda(\bm{q},\omega+\nu,-\omega)
=
\left[
1
-
\frac{i}{2\tau_{\rm imp}}
\frac{1}{2\bar{q}}\,
\log\!\left(
\frac{\omega+\nu/2+i\tilde{\gamma}/2+\bar{q}}
{\omega+\nu/2+i\tilde{\gamma}/2-\bar{q}}
\right)
\right]^{-1}.
\label{eq.vertex}	
\end{equation}
\end{widetext}
The superconducting phase transition temperature $T_{\rm env}^{\rm c}$, as well as the pair chemical potential $\mu_{\rm pair}$, are determined by solving the Thouless criterion in Eqs.~\eqref{eq.Thouless.Re} and \eqref{eq.Thouless.Im}, together with Eqs.~\eqref{eq.PiR} and \eqref{eq.vertex}. The momentum $\bm{Q}$ in Eqs.~\eqref{eq.Thouless.Re} and \eqref{eq.Thouless.Im} is chosen so as to maximize the critical temperature $T_{\rm env}^{\rm c}$. The solution with $\bm{Q}=0$ corresponds to the NBCS state, whereas a non-zero $\bm{Q}$ signals a transition to the spatially non-uniform NFFLO state.

Since Eqs.~\eqref{eq.PiR} and \eqref{eq.vertex} do not depend on the electrostatic potential $\varphi$, the equations determining $T_{\rm env}^{\rm c}$ and $\mu_{\rm pair}$ are closed within Eqs.~\eqref{eq.Thouless.Re} and \eqref{eq.Thouless.Im}. Once $T_{\rm env}^{\rm c}$ and $\mu_{\rm pair}$ are obtained, the electrostatic potential $\varphi$ is determined from the charge neutrality condition in the superconductor~\cite{Kopnin2001}. The corresponding charge density is obtained from the Keldysh Green’s function as~\cite{Rammer2007, Chandrasekhar2004}
\begin{align}
\delta \rho_\sigma
&=
\frac{-ie}{2}\sum_{\bm{k}}\int_{-\infty}^\infty
\frac{d\omega}{2\pi}\,
G^{\rm K}_\sigma(\bm{k},\omega)
\notag\\
&=
-\frac{eN(0)}{2}\int_{-\infty}^\infty d\omega
\big[1-2f_{\rm neq}(\omega)\big]
-
e^2N(0)\varphi.
\label{eq.del.rho.sigma}
\end{align}
Here, we use $\delta\rho_\sigma$ instead of $\rho_\sigma$ to emphasize that Eq.~\eqref{eq.del.rho.sigma} does not involve the equilibrium contributions. In deriving the second line of Eq.~\eqref{eq.del.rho.sigma}, we have used particle–hole symmetry and have extended the $\xi_{\bm{k}}$-integration to $(-\infty,\infty)$. The electrostatic potential enters as a uniform shift of the single-particle energies, $\xi_{\bm k}\to\xi_{\bm k}+e\varphi$, and the resulting shift yields the second term~\cite{Rammer2007}. Since charge neutrality must be satisfied in a metallic superconductor at low temperatures~\cite{Eschrig2009, Heikkila2013}, we impose the condition $\delta\rho_\sigma=0$, which yields
\begin{equation}
e\varphi
=
-\frac{1}{2}
\int_{-\infty}^\infty d\omega
\big[1-2f_{\rm neq}(\omega)\big].
\label{eq.varphi}
\end{equation}

\subsection{Nonequilibrium BCS theory \label{sec.formalism.NBCS}}

We now develop a nonequilibrium mean-field theory for the NBCS state. We focus on a nonequilibrium steady state characterized by a BCS-type superconducting order parameter,
\begin{equation}
\Delta(t)
=
U\sum_{\bm{k}}
\braket{a_{-\bm{k},\down}(t) a_{\bm{k},\up}(t)}
=
\Delta e^{-2i\mu_{\rm pair} t}.
\label{eq.OP}
\end{equation}
Here, $\mu_{\rm pair}$ denotes the electrochemical potential of Cooper pairs, which also appears in the Thouless criterion in Eqs.~\eqref{eq.Thouless.Re} and \eqref{eq.Thouless.Im}. As schematically illustrated in Fig.~\ref{Fig.energy}(a), the pair chemical potential $\mu_{\rm pair}$ lies within the bias window $[0,eV]$. When energies are measured relative to the Fermi level of the lead~2, a non-zero pair chemical potential $\mu_{\rm pair}\neq0$ introduces the time-dependent phase factor $e^{-2i\mu_{\rm pair} t}$ into the order parameter $\Delta(t)$~\cite{Josephson1962, Kopnin2001}. To eliminate this trivial time dependence and simplify the subsequent analysis, we conveniently shift the energy reference to $\mu_{\rm pair}$, as shown in Fig.~\ref{Fig.energy}(b). We note that $\mu_{\rm pair}$ is self-consistently determined from the steady-state condition, which will later be given in Eq.~\eqref{eq.NESS}.

To describe the nonequilibrium superconducting state, we extend the Keldysh Green's function introduced in Eq.~\eqref{eq.Ghat}
to the Nambu $\otimes$ spin space,
\begin{equation}
\hat{\bm{G}}(\bm{k},\omega)
=
\begin{pmatrix}
\bm{G}^{\rm R}(\bm{k},\omega) &
\bm{G}^{\rm K}(\bm{k},\omega) \\[4pt]
0 &
\bm{G}^{\rm A}(\bm{k},\omega)
\end{pmatrix},
\label{eq.Nambu.Keldysh.G}
\end{equation}
where the retarded, advanced, and Keldysh components are defined as
\begin{subequations}
\begin{align}
\bm{G}^{\rm R}(\bm{k},\omega)
&=
-i \int_{-\infty}^\infty dt\,
\Theta(t)\,
\braket{[\bm{\Psi}_{\bm{k}}(t), \bm{\Psi}^\dagger_{\bm{k}}(0)]_+},
\\
\bm{G}^{\rm A}(\bm{k},\omega)
&=
\big[\bm{G}^{\rm R}(\bm{k},\omega)\big]^\dagger,
\\
\bm{G}^{\rm K}(\bm{k},\omega)
&=
-i \int_{-\infty}^\infty dt\,
\braket{[\bm{\Psi}_{\bm{k}}(t), \bm{\Psi}^\dagger_{\bm{k}}(0)]_-},
\end{align}
\label{eq.def.Nambu.Green}
\end{subequations}
with
\begin{equation}
\bm{\Psi}_{\bm{k}}^\dagger
=
\begin{pmatrix}
a^\dagger_{\bm{k},\up} &
a^\dagger_{\bm{k},\down} &
a_{-\bm{k},\up} &
a_{-\bm{k},\down}
\end{pmatrix}
\label{eq.Namnbu.4}
\end{equation}
being the four-component Nambu field. The Nambu-Keldysh Green's function $\hat{\bm{G}}$ in Eq.~\eqref{eq.Nambu.Keldysh.G} obeys the Dyson equation,
\begin{equation}
\hat{\bm{G}}(\bm{k},\omega)
=
\hat{\bm{G}}_0(\bm{k},\omega)
+
\hat{\bm{G}}_0(\bm{k},\omega)
\hat{\bm{\Sigma}}(\bm{k},\omega)
\hat{\bm{G}}(\bm{k},\omega).
\label{eq.Dyson.Nambu}
\end{equation}
Here, $\hat{\bm{G}}_0$ has the same matrix structure as $\hat{\bm{G}}$ in Eq.~\eqref{eq.Nambu.Keldysh.G}, where each component has the form,
\begin{subequations}
\begin{align}
\bm{G}^{\rm R}_0(\bm{k},\omega)
&=
\frac{1}{[\omega +i\delta]\, \big(\tau_0\otimes \sigma_0\big) 
-[\xi_{\bm{k}}+e\varphi]\, \big(\tau_3\otimes \sigma_0\big)},
\label{eq.G0R.Nambu}
\\
\bm{G}^{\rm A}_0(\bm{k},\omega)
&=
\big[\bm{G}^{\rm R}_0(\bm{k},\omega)\big]^\dagger,
\\
\bm{G}^{\rm K}_0(\bm{k},\omega)
&=
\big[\bm{G}^{\rm R}_0(\bm{k},\omega) -\bm{G}^{\rm A}_0(\bm{k},\omega)\big]\big[1-2f_{\rm ini}(\omega)\big].
\end{align}
\end{subequations}
This $\hat{\bm{G}}_0$ corresponds to the Nambu representation of the bare Green's function in Eq.~\eqref{eq.G0}. In Eq.~\eqref{eq.G0R.Nambu}, $\tau_0$ and $\sigma_0$ denote unit matrices, and $\tau_{i=1,2,3}$ and $\sigma_{\nu=x,y,z}$ are Pauli matrices acting on particle–hole space and spin space, respectively. The direct products $\tau_i\otimes\sigma_\nu$ act on the four-component Nambu $\otimes$ spin space.

The self-energy in Eq.~\eqref{eq.Dyson.Nambu} consists of four contributions,
\begin{align}
\hat{\bm{\Sigma}}(\bm{k},\omega)
&=
\hat{\bm{\Sigma}}_{\rm int}(\bm{k},\omega)
+
\hat{\bm{\Sigma}}_{\rm lead}(\bm{k},\omega)
\notag\\
&\hspace{1.5cm}
+
\hat{\bm{\Sigma}}_{\rm imp}(\bm{k},\omega)
+
\hat{\bm{\Sigma}}_{\rm mag}(\bm{k},\omega).
\label{eq.self.Nambu}	
\end{align}
Among these, $\hat{\bm{\Sigma}}_{\rm int}$ describes effects of the pairing interaction, which is given by, within the mean-field BCS approximation~\cite{Kawamura2022},
\begin{equation}
\hat{\bm{\Sigma}}_{\rm int}(\bm{k},\omega)
=
\begin{pmatrix}
\Delta\, \big(\tau_2\otimes \sigma_y\big) & 0 \\[4pt]
0 & \Delta\, \big(\tau_2\otimes \sigma_y\big)
\end{pmatrix}.
\label{eq.self.int}
\end{equation}
Here, $\Delta$ is the steady-state superconducting order parameter in Eq.~\eqref{eq.OP}. We note that the order parameter is time independent in the present formulation, because energies are measured from the pair chemical potential $\mu_{\rm pair}$, as shown in Fig.~\ref{Fig.energy}(b).

Effects of the normal-metal leads are incorporated through
$\hat{\bm{\Sigma}}_{\rm lead}$ in Eq.~\eqref{eq.self.Nambu}, which is given by~\cite{Kawamura2022}
\begin{subequations}
\begin{align}
&
\bm{\Sigma}^{\rm R}_{\rm lead}(\bm{k},\omega)
=
\big[\bm{\Sigma}^{\rm A}_{\rm lead}(\bm{k},\omega)\big]^\dagger
=
-\frac{i\gamma}{2}\, \big(\tau_0\otimes \sigma_0\big),
\\[4pt]
&
\bm{\Sigma}^{\rm K}_{\rm lead}(\bm{k},\omega)
=
-i \sum_{\alpha=1,2} \gamma_\alpha
\notag\\
&
\times
\scalebox{0.8}{$\displaystyle
\begin{pmatrix}
\big[1-2f(\omega +\mu_{\rm pair} -eV\delta_{\alpha,1})\big] \sigma_0 
& 0 \\[4pt]
0 & 
\big[1-2f(\omega -\mu_{\rm pair} +eV\delta_{\alpha,1})\big] \sigma_0
\end{pmatrix}$}.
\label{eq.self.lead.K.Nambu}
\end{align}
\end{subequations}
This expression corresponds to the Nambu representation of the lead self-energy in Eq.~\eqref{eq.self.lead}. In the Keldysh component $\bm{\Sigma}^{\rm K}_{\rm lead}$, the arguments of the distribution functions $f(\omega)$ are shifted by $\mu_{\rm pair}$ because of the present choice of the energy reference.

The remaining self-energy contributions, $\hat{\bm{\Sigma}}_{\rm imp}$ and $\hat{\bm{\Sigma}}_{\rm mag}$, describe effects of elastic scattering from nonmagnetic and magnetic impurities, respectively. Within the framework of the self-consistent Born approximation, these  are evaluated as~\cite{Maki1969, Kopnin2001}
\begin{align}
&
\bm{\Sigma}^{{\rm X}={\rm R},{\rm A},{\rm K}}_{\rm imp}(\bm{k},\omega)
\notag\\
&\hspace{0.5cm}
=
n_{\rm imp} u_{\rm imp}^2
\sum_{\bm{k}}
\big(\tau_3 \otimes \sigma_0\big)\,
\bm{G}^{\rm X}(\bm{k},\omega)\,
\big(\tau_3 \otimes \sigma_0\big),
\label{eq.self.imp.Nambu}
\\
&
\bm{\Sigma}^{\rm X}_{\rm mag}(\bm{k},\omega)
\notag\\
&\hspace{0.5cm}
=
n_{\rm mag} u_{\rm mag}^2\,
\frac{S(S+1)}{3}
\sum_{\nu=x,y,z}\sum_{\bm{k}}
\bm{\alpha}_\nu\,
\bm{G}^{\rm X}(\bm{k},\omega)\,
\bm{\alpha}_\nu,
\label{eq.self.mag.Nambu}
\end{align}
where $n_{\rm mag}=N_{\rm mag}/V$ and
\begin{equation}
\bm{\alpha}_{\nu=x,y,z}
=
\begin{pmatrix}
\sigma_\nu & 0 \\[4pt]
0 & -\sigma^{\rm t}_\nu
\end{pmatrix}.
\end{equation}
In deriving Eq.~\eqref{eq.self.mag.Nambu}, we have assumed that the impurity spins are randomly oriented and have taken the average over the impurity spin orientations.

The dressed Green’s function $\hat{\bm{G}}$ is obtained by substituting the self-energies in Eqs.~\eqref{eq.self.int}-\eqref{eq.self.mag.Nambu} into the Dyson equation~\eqref{eq.Dyson.Nambu}. Because $\hat{\bm{\Sigma}}_{\rm imp}$ and $\hat{\bm{\Sigma}}_{\rm mag}$ depend explicitly on $\hat{\bm{G}}$, the Dyson equation must be solved self-consistently. For this purpose, we employ the standard technique widely used in the study of impurity effects on thermal-equilibrium superconductivity~\cite{Maki1969}: We first focus on the retarded component of the Dyson equation~\eqref{eq.Dyson.Nambu}, which can be formally solved as
\begin{align}
&
\bm{G}^{\rm R}(\bm{k}, \omega) 
\notag\\[4pt]
&
=
\Big[
\omega_+\, \big(\tau_0\otimes \sigma_0\big) 
-[\xi_{\bm{k}}+e\varphi]\, \big(\tau_3\otimes \sigma_0\big) 
\notag\\
&\hspace{1.5cm}
-\Delta\, \big(\tau_2\otimes \sigma_y\big) 
-\bm{\Sigma}^{\rm R}_{\rm imp}(\omega) 
-\bm{\Sigma}^{\rm R}_{\rm mag}(\omega)
\Big]^{-1}
\notag\\[4pt]
&=
\frac{
\tilde{\omega}_+\, \big(\tau_0\otimes \sigma_0\big)+
[\xi_{\bm{k}}+e\varphi]\, \big(\tau_3\otimes \sigma_0\big)+
\tilde{\Delta}(\omega)\, \big(\tau_2\otimes \sigma_y\big)
}{\tilde{\omega}_+^2 -[\xi_{\bm{k}}+e\varphi]^2 -\tilde{\Delta}(\omega)^2},
\label{eq.dressed.GR}
\end{align}

with $\omega_+ = \omega + i\gamma/2$. Here, we have introduced the renormalized frequency $\tilde{\omega}_+$ and the renormalized gap function $\tilde{\Delta}(\omega)$, defined through
\begin{align}
&
\bm{\Sigma}^{\rm R}_{\rm imp}(\omega)+ 
\bm{\Sigma}^{\rm R}_{\rm mag}(\omega)
\notag\\[4pt]
&\hspace{0.2cm}
=
\big[\omega_+ - \tilde{\omega}_+\big]\,
\big(\tau_0\otimes \sigma_0\big)
-
\big[\Delta -\tilde{\Delta}(\omega) \big]\,
\big(\tau_2\otimes \sigma_y\big).
\label{eq.def.til}	
\end{align}
Substitution of Eq.~\eqref{eq.dressed.GR} into Eqs.~\eqref{eq.self.imp.Nambu} and \eqref{eq.self.mag.Nambu}, which is followed by the $\bm{k}$ summation, gives
\begin{subequations}
\begin{align}
&
\bm{\Sigma}^{\rm R}_{\rm imp}(\omega)=
-\frac{1}{2\tau_{\rm imp}}\,
\frac{\tilde{\omega}_+ -\tilde{\Delta}(\omega)\,\big(\tau_2 \otimes \sigma_y\big)}{\sqrt{\tilde{\Delta}(\omega)^2 -\tilde{\omega}_+^2}}
,\\[4pt]
&
\bm{\Sigma}^{\rm R}_{\rm mag}(\omega)
=
-\frac{1}{2\tau_{\rm mag}}\,
\frac{\tilde{\omega}_+ +\tilde{\Delta}(\omega)\,\big(\tau_2 \otimes \sigma_y\big)}{\sqrt{\tilde{\Delta}(\omega)^2 -\tilde{\omega}_+^2}}.
\label{eq.self.SigR.mag.Nambu}
\end{align}
\label{eq.til.self.Nambu}
\end{subequations}
Here, the elastic scattering rate $\tau_{\rm imp}$ due to nonmagnetic impurities is given in Eq.~\eqref{eq.def.tau.imp}. In Eq.~\eqref{eq.self.SigR.mag.Nambu},
\begin{equation}
\frac{1}{\tau_{\rm mag}}=
2\pi n_{\rm mag} u_{\rm mag}^2 S(S+1) N(0)
\end{equation}
is the elastic scattering rate due to magnetic impurities. Substituting Eq.~\eqref{eq.til.self.Nambu} into Eq.~\eqref{eq.def.til}, one reaches the following coupled self-consistent equations for $\tilde{\omega}_+$ and $\tilde{\Delta}(\omega)$:
\begin{subequations}
\begin{align}
&
\tilde{\omega}_+ = 
\omega_+ +\frac{1}{2}\left[\frac{1}{\tau_{\rm imp}} + \frac{1}{\tau_{\rm mag}} \right]\, \frac{\tilde{\omega}_+}{\sqrt{\tilde{\Delta}(\omega)^2 -\tilde{\omega}_+^2}}
,\\
&
\tilde{\Delta}(\omega) =
\Delta +\frac{1}{2}\left[\frac{1}{\tau_{\rm imp}} - \frac{1}{\tau_{\rm mag }} \right]\, \frac{\tilde{\Delta}(\omega)}{\sqrt{\tilde{\Delta}(\omega)^2 -\tilde{\omega}_+^2}}.	
\end{align}
\label{eq.til}
\end{subequations}
We numerically solve Eq.~\eqref{eq.til}, to self-consistently determine $\tilde{\omega}_+$ and $\tilde{\Delta}(\omega)$. We then substitute the resulting solutions into Eq.~\eqref{eq.dressed.GR}. This procedure allows an efficient evaluation of the dressed retarded Green’s function $\bm{G}^{\rm R}$. Once the retarded Green’s function is obtained, the Keldysh component $\bm{G}^{\rm K}$ can be evaluated from the Keldysh sector of the Dyson equation~\eqref{eq.Dyson.Nambu}, given by~\cite{Rammer2007}
\begin{align}
&
\bm{G}^{\rm K}(\bm{k},\omega)
\notag\\[4pt]
&
=
\bm{G}^{\rm R}(\bm{k},\omega)
\big[
\bm{\Sigma}^{\rm K}_{\rm lead}(\omega) +
\bm{\Sigma}^{\rm K}_{\rm imp}(\omega) +
\bm{\Sigma}^{\rm K}_{\rm mag}(\omega)
\big]
\bm{G}^{\rm A}(\bm{k},\omega).	
\end{align}
We numerically evaluate $\bm{\Sigma}^{\rm K}_{\rm imp}$ and $\bm{\Sigma}^{\rm K}_{\rm mag}$, to determine $\bm{G}^{\rm K}$ in a fully self-consistent manner.

Physical quantities can be determined from the dressed Green's function $\hat{\bm{G}}$. The superconducting order parameter $\Delta$ is obtained from the gap equation,
\begin{equation}
\Delta
=
-\frac{i U N(0)}{2}\int_{-\omega_{\rm D}}^{\omega_{\rm D}}
\frac{d\omega}{2\pi}\, {\rm Tr}\Big[\big(\tau_- \otimes \sigma_-\big)\, \bm{g}^{\rm K}(\omega)\Big],
\label{eq.noneq.gap}
\end{equation}
where $\omega_{\rm D}$ denotes the Debye frequency cutoff, $\tau_-\equiv [\tau_1+i\tau_2]/2$, and $\sigma_-\equiv [\sigma_1+i\sigma_2]/2$. In Eq.~\eqref{eq.noneq.gap},
\begin{equation}
\bm{g}^{\rm X}(\omega)
\equiv
\int d\xi \, \bm{G}^{\rm X}(\bm{k},\omega)
\label{eq.g.qc}
\end{equation}
is commonly referred to as the quasiclassical Green's function in the literature~\cite{Rammer2007, Kopnin2001}.

The pair chemical potential $\mu_{\rm pair}$ is determined from the steady-state condition $I_1+I_2=0$, where
\begin{align}
&
\scalebox{0.95}{$\displaystyle
I_\alpha=
\frac{i e \gamma_\alpha N(0)}{2}
\int_{-\infty}^{\infty}
\frac{d\omega}{2\pi}\,
{\rm Tr}\bigg[
\big(\tau_{\rm p}\otimes \sigma_0\big)$}
\notag\\
&
\scalebox{0.95}{$\displaystyle
\times
\Big[
-\big[1-2f(\omega+\mu_{\rm pair}-\mu_\alpha)\big]
\big[\bm{g}^{\rm R}(\omega)-\bm{g}^{\rm A}(\omega)\big]
+
\bm{g}^{\rm K}(\omega)
\Big]
\bigg]$},
\end{align}
is the current flowing from the lead~$\alpha$ into the superconductor~\cite{Jauho1994, Meir1992}, with $\tau_{\rm p}\equiv [1+\tau_3]/2$ being the projector onto the particle sector. Accordingly, the steady-state condition can be expressed as follows:
\begin{align}
&
\scalebox{0.98}{$\displaystyle
0=
\int_{-\infty}^{\infty} d\omega\,
{\rm Tr}\bigg[
\big(\tau_{\rm p}\otimes \sigma_0\big)$}
\notag\\
&
\scalebox{0.98}{$\displaystyle
\times
\Big[
-\big[1-2f_{\rm neq}(\omega+\mu_{\rm pair})\big]
\big[\bm{g}^{\rm R}(\omega)-\bm{g}^{\rm A}(\omega)\big]
+
\bm{g}^{\rm K}(\omega)
\Big]
\bigg]$}.
\label{eq.NESS}
\end{align}
Here, $f_{\rm neq}(\omega)$ is given in Eq.~\eqref{eq.fneq.def}. We determine the order parameter $\Delta$ and the pair chemical potential $\mu_{\rm pair}$, by self-consistently solving Eqs.~\eqref{eq.noneq.gap} and \eqref{eq.NESS}. In these equations, the retarded and advanced components $\bm{g}^{\rm R, A}$ are given by
\begin{equation} 
\bm{g}^{\rm R}(\omega)= \big[\bm{g}^{\rm A}(\omega)\big]^\dagger= -\pi\, \frac{\tilde{\omega}_+ +\tilde{\Delta}(\omega)\,\big(\tau_2 \otimes \sigma_y\big)}{\sqrt{\tilde{\Delta}(\omega)^2 -\tilde{\omega}_+^2}}. 
\label{eq.gr.qc} 
\end{equation}
The Keldysh component $\bm{g}^{\rm K}$ is numerically evaluated from Eq.~\eqref{eq.g.qc}. We note that, in the limit $\Delta \to +0$, Eqs.~\eqref{eq.noneq.gap} and \eqref{eq.NESS} reduce to the real and imaginary parts of the Thouless criterion in Eqs.~\eqref{eq.Thouless.Re} and \eqref{eq.Thouless.Im}, respectively.

As in Sec.~\ref{sec.Thouless}, the electrostatic potential $\varphi$ is determined from the charge neutrality condition $\delta\rho=0$: Since the charge density is given by~\cite{Rammer2007, Chandrasekhar2004}
\begin{align}
\delta\rho 
&=
-\frac{ie}{4}\sum_{\bm{k}}\int_{-\infty}^\infty \frac{d\omega}{2\pi}\,
{\rm Tr}\Big[
\big(\tau_3 \otimes \sigma_0\big)\,
\bm{G}^{\rm K}(\bm{k},\omega)\Big]
\notag\\[4pt]
&=
2eN(0)
\left[
e\varphi
+
\frac{i}{8}
\int_{-\infty}^{\infty}
\frac{d\omega}{2\pi}\,
{\rm Tr}\Big[
\big(\tau_3 \otimes \sigma_0\big)\,
\bm{g}^{\rm K}(\omega)
\Big]
\right],
\label{eq.rho.noneq.Nambu2}
\end{align}
the electrostatic potential is determined as
\begin{equation}
e\varphi
=
-\frac{i}{8}
\int_{-\infty}^{\infty}
\frac{d\omega}{2\pi}\,
{\rm Tr}\Big[
\big(\tau_3 \otimes \sigma_0\big)\,
\bm{g}^{\rm K}(\omega)
\Big].
\label{eq.varphi.Nambu}
\end{equation}
It is then convenient to introduce the following quasiparticle electrochemical potential~\cite{Kopnin2001}:
\begin{equation}
\mu_{\rm qp} = e\varphi +\frac{\delta \rho}{N(0)}.
\label{eq.muqp.def}
\end{equation}
Under the charge neutrality condition $\delta\rho=0$, one then finds
\begin{equation}
\mu_{\rm qp}=e\varphi.
\label{eq.muqp}
\end{equation}
In the thermal equilibrium state, quasiparticles and Cooper pairs are equilibrium with each other, which requires their chemical potentials to balance, $\mu_{\rm pair}=\mu_{\rm qp}$ ($=e\varphi$). In the nonequilibrium superconducting state, by contrast, this condition is lifted, and the two chemical potentials need not coincide with each other, as will be demonstrated later.

\section{Effects of Lead-Coupling Asymmetry on the NFFLO State \label{sec.FFLO.asy}}

In this section, we examine how asymmetry ($P_{\rm lead}\neq 0$) between the two superconductor–lead couplings affects the NFFLO state.

\begin{figure}[t]
\centering
\includegraphics[width=\columnwidth]{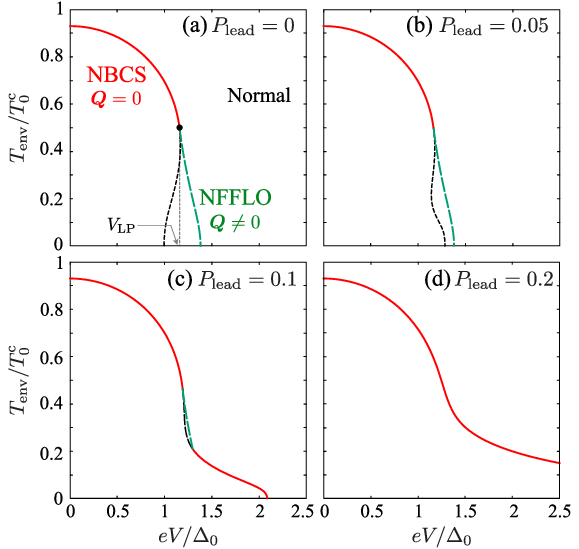}
\caption{
Calculated superconducting phase transition temperature $T_{\rm env}^{\rm c}$ as a function of bias voltage $V$ in the presence of asymmetry ($P_{\rm lead}>0$) between the two superconductor–lead couplings. The solid and dashed lines represent the transitions to the NBCS ($\bm{Q}=0$) and NFFLO ($\bm{Q}\neq 0$) states, respectively. We set the total lead-coupling strength to $\gamma/\Delta_0=0.1$, and the clean limit $\tau_{\rm imp}=\infty$ is considered. $\Delta_0$ and $T_0^{\rm c}$ denote the superconducting order parameter and critical temperature in the case of thermal-equilibrium superconductivity without coupling to the leads, respectively.
}
\label{Fig.FFLO.asy}
\end{figure}

Figure~\ref{Fig.FFLO.asy} shows the superconducting phase transition temperature obtained from the Thouless criterion in Eqs.~\eqref{eq.Thouless.Re} and \eqref{eq.Thouless.Im} in the case where $P_{\rm lead}\neq 0$ ($\gamma_1< \gamma_2$ and $\gamma/\Delta_0 = 0.1$). Since we focus on how the coupling asymmetry affects the NFFLO state, we consider the clean system by setting $\tau_{\rm imp} \to \infty$, for simplicity. In the symmetric case ($P_{\rm lead}=0$) shown in Fig.~\ref{Fig.FFLO.asy}(a), the NFFLO phase transition occurs at $T_{\rm env}/T_0^{\rm c} \lesssim 0.5$. However, as the coupling asymmetry $P_{\rm lead}$ is increased, one sees from Figs.~\ref{Fig.FFLO.asy}(b) and (c) that the NFFLO region is progressively suppressed and eventually vanishes when $P_{\rm lead}=0.2$, as shown in Fig.~\ref{Fig.FFLO.asy}(d). Thus, coupling asymmetry is found to be detrimental to the NFFLO state.

\begin{figure}[t]
\centering
\includegraphics[width=\columnwidth]{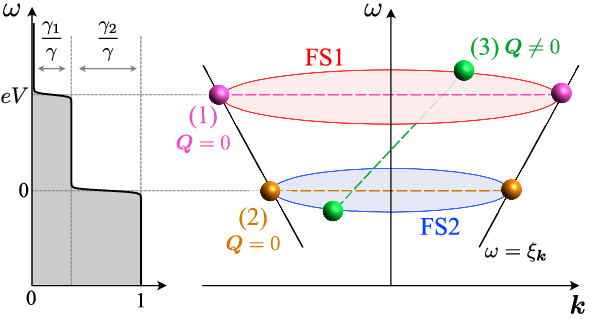}
\caption{Cooper pairings associated with the effective Fermi surfaces (FS1 and FS2) induced by the nonequilibrium electron distribution $f_{\rm neq}(\omega)$. In the intra-surface pairing channels (1) and (2), electrons on the same effective Fermi surface form Cooper pairs with zero center-of-mass momentum, resulting in the uniform NBCS state. By contrast, the inter-surface pairing channel (3) yields Cooper pairs with a non-zero center-of-mass momentum, leading to the NFFLO state. The relative electron populations on FS1 and FS2 are approximately given by the lead-coupling ratio $\gamma_1:\gamma_2$. 
}
\label{Fig.FS}
\end{figure}

The suppression of the NFFLO state when $P_{\rm lead}\neq 0$ can be understood in terms of the effective ``Fermi surfaces" induced by the nonequilibrium electron distribution $f_{\rm neq}(\omega)$: As mentioned in Sec.~\ref{sec.Intro}, the two-step structure of $f_{\rm neq}(\omega)$ gives rise to two effective Fermi surfaces, denoted as FS1 and FS2 (see Fig.~\ref{Fig.FS}). These effective Fermi surfaces open multiple Cooper-pairing channels, schematically shown in Fig.~\ref{Fig.FS}. In pairing channels (1) and (2), where electrons on the same effective Fermi surface form Cooper pairs, the center-of-mass momentum of the pairs is zero, resulting in the uniform NBCS state. By contrast, the inter-Fermi-surface pairing channel (3) yields Cooper pairs with nonzero center-of-mass momentum, leading to the NFFLO state.

A key point is that electrons on FS1 and FS2 are characterized by different effective ``Fermi energies". As a result, the pairing channels (1) and (2) are associated with different pair chemical potentials. Since a superconducting condensate is characterized by a single pair chemical potential $\mu_{\rm pair}$, these two pairing channels cannot be simultaneously realized. Once a Cooper instability is triggered by FS1, electrons on FS2 do not contribute to Cooper pairing, and vice versa. By contrast, the inter-surface pairing channel (3) involves electrons on both FS1 and FS2 and thus gains a larger condensation energy than the NBCS-type pairings, where only a single effective Fermi surface participates. This explains why, as shown in Figs.~\ref{Fig.FFLO.asy}(a) and (b), the NFFLO instability dominates at low temperatures when the lead-coupling asymmetry $P_{\rm lead}$ is small.

The situation qualitatively changes when the couplings to the two leads ($\alpha=1,2$) become strongly asymmetric. In this case, the nonequilibrium electron distribution is given by Eq.~\eqref{eq.fneq.def}, and, as illustrated in Fig.~\ref{Fig.FS}, the electron population on FS1 becomes smaller than that on FS2. As a result, the inter-surface pairing channel (3) is limited by the reduced population on FS1, while the excess electrons on FS2 do not contribute to Cooper pairing. Consequently, the advantage of the condensation energy of the NFFLO state in the nearly symmetric case is lost. For sufficiently strong asymmetry, the NBCS-type intra-surface pairing in channel (2) becomes dominant, and the Cooper instability is driven primarily by FS2. This explains the vanishing NFFLO instability in Fig.~\ref{Fig.FFLO.asy}(d).

\begin{figure}[t]
\centering
\includegraphics[width=\columnwidth]{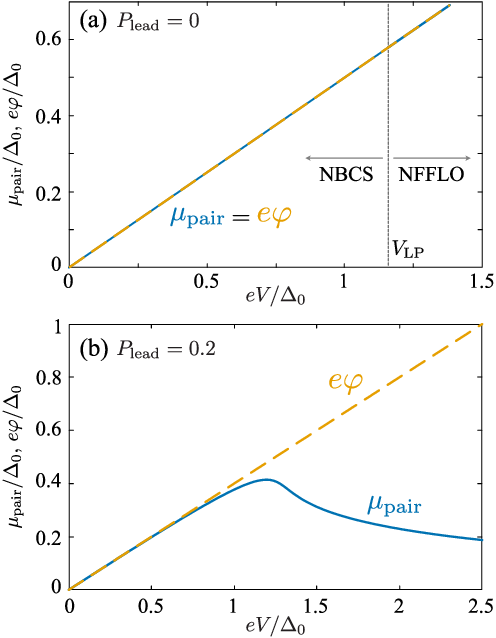}
\caption{
Pair chemical potential $\mu_{\rm pair}$ (solid line) and electrostatic potential $\varphi$ (dashed line) along the superconducting phase boundary as functions of the bias voltage $V$. (a) $P_{\rm lead}=0$ (symmetric case). (b) $P_{\rm lead}=0.2$ (asymmetric case). In panel (a), the system is in the NFFLO (NBCS) state when $V > V_{\rm LP}$ ($V < V_{\rm LP}$), where $V_{\rm LP}$ is the voltage of the Lifshitz point shown in Fig.~\ref{Fig.FFLO.asy}(a).
}
\label{Fig.mu.asy}
\end{figure}

Further insight into the effects of asymmetry between the two superconductor–lead couplings is obtained from the pair chemical potential $\mu_{\rm pair}$: Figures~\ref{Fig.mu.asy}(a) and (b) show $\mu_{\rm pair}$ and $\varphi$ as functions of the bias voltage $V$, evaluated along the superconducting phase boundaries shown in Figs.~\ref{Fig.FFLO.asy}(a) and (d), respectively. As illustrated in Fig.~\ref{Fig.energy}(a), all energies are measured from the Fermi level of the lead 2. In Fig.~\ref{Fig.mu.asy}(a), the region $V>V_{\rm LP}$ corresponds to the NFFLO phase, where $V_{\rm LP}$ denotes the voltage of the Lifshitz point seen in Fig.~\ref{Fig.FFLO.asy}(a).

The pair chemical potential $\mu_{\rm pair}$ provides information on which effective Fermi surface(s) predominantly drive the Cooper instability. In the NFFLO state, $\mu_{\rm pair}$ follows the linear relation $\mu_{\rm pair}=eV/2$, as seen in Fig.~\ref{Fig.mu.asy}(a). This reflects the inter-surface nature of Cooper pairing in the NFFLO state, where electrons near both FS1 and FS2 participate in pairing channel (3) in Fig.~\ref{Fig.FS}. The pair chemical potential $\mu_{\rm pair}$ thus lies at the average of the effective ``Fermi energies" of FS1 and FS2, which are located at $eV$ and $0$, respectively.

In contrast, when the asymmetry of the superconductor–lead couplings becomes sufficiently pronounced, $\mu_{\rm pair}$ exhibits a nonmonotonic bias-voltage dependence, as shown in Fig.~\ref{Fig.mu.asy}(b): It initially increases with increasing bias voltage but starts to decrease around $eV/\Delta_0 \simeq 1.2$ (which corresponds to $T_{\rm env}/T_0^{\rm c} \simeq 0.6$). This turning point marks the onset of the regime $eV \gtrsim 2T_{\rm env}$, where the bias voltage exceeds the thermal broadening of the lead distribution functions. In this regime, the two-step structure of the nonequilibrium electron distribution becomes well developed, activating the two effective Fermi surfaces. When the coupling asymmetry is strong, the intra-surface pairing channel (2) on FS2 then becomes dominant, pulling $\mu_{\rm pair}$ down toward the Fermi level of the lead 2 (i.e., $\mu_{\rm pair}=0$).

We note that the electrostatic potential $\varphi$ exhibits a qualitatively different behavior from the pair chemical potential $\mu_{\rm pair}$. Irrespective of whether the superconductor–lead coupling is symmetric or asymmetric, $\varphi$ increases monotonically with increasing the bias voltage, as shown in Fig.~\ref{Fig.mu.asy}. This behavior can be understood from the following simple estimate: At sufficiently low temperatures, the Fermi–Dirac distribution can be approximated by the step function as $f(\omega)\simeq 1-\Theta(\omega)$. The nonequilibrium distribution function in Eq.~\eqref{eq.fneq.def} is then approximated as
\begin{equation}
f_{\rm neq}(\omega) \simeq
1
-\frac{\gamma_1\Theta(\omega-eV) +\gamma_2\Theta(\omega)}{\gamma_1+\gamma_2}.
\end{equation}
Substituting this into Eq.~\eqref{eq.varphi}, one can estimate the electrostatic potential as
\begin{equation}
\varphi \simeq \frac{\gamma_1}{\gamma_1+\gamma_2}, V.
\end{equation}
Thus, the electrostatic potential is found to increase monotonically in proportion to the applied bias voltage, which is independent of how $\mu_{\rm pair}$ behaves.

As shown in Fig.~\ref{Fig.mu.asy}(b), when the couplings are asymmetric ($P_{\rm lead}\neq 0$), a clear separation between $\mu_{\rm pair}$ and $e\varphi$ is seen at large bias voltages. Since the quasiparticle chemical potential is given by $\mu_{\rm qp} = e\varphi$ [see Eq.~\eqref{eq.muqp}], this separation indicates the presence of a chemical-potential imbalance between quasiparticles and the condensate. The microscopic origin of this potential imbalance $\mu_{\rm pair}\neq \mu_{\rm qp}$ is discussed in detail in Sec.~\ref{sec.NBCS.imbalance} from the perspective of the so-called charge imbalance phenomenon.

In the symmetric case ($P_{\rm lead}=0$), Fig.~\ref{Fig.mu.asy}(a) shows that the chemical potentials are always balanced, that is, $\mu_{\rm pair}=\mu_{\rm qp}(=e\varphi)$. This indicates that charge imbalance is absent in a symmetric NSN junction. While this has been pointed out in previous work for BCS-type {\it uniform} nonequilibrium superconductivity~\cite{Takahashi1999}, Fig.~\ref{Fig.mu.asy}(a) demonstrates that the same conclusion holds even in the presence of the {\it nonuniform} NFFLO instability.

\section{Effects of Lead-Coupling Asymmetry on the NBCS State \label{sec.NBCS}}

In this section, we examine the nonequilibrium properties of the NBCS state in the presence of asymmetry between the two superconductor–lead couplings. In Sec.~\ref{sec.NBCS.steady}, we present steady-state solutions to show that, when $P_{\rm lead}\neq 0$, the NBCS state can be classified into two types: one is accompanied by the chemical-potential imbalance ($\mu_{\rm pair}\neq\mu_{\rm qp}$), and the other is not. In Sec.~\ref{sec.NBCS.imbalance}, we elucidate the microscopic origin of the chemical-potential imbalance based on Tinkham’s branch-imbalance picture.

\subsection{Steady-state solutions of the NBCS state \label{sec.NBCS.steady}}

\begin{figure}[t]
\centering
\includegraphics[width=\columnwidth]{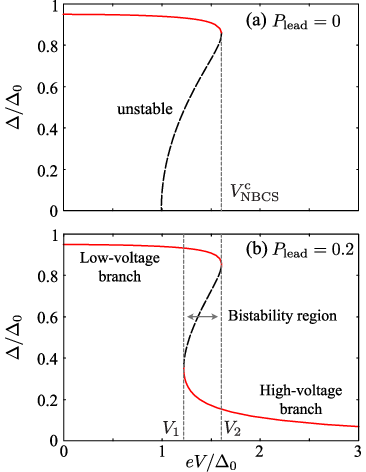}
\caption{
Calculated steady-state solutions of the NBCS state. (a) Symmetric coupling case ($P_{\rm lead}=0$). (b) Asymmetric coupling case ($P_{\rm lead}=0.2$). Solid (dashed) lines indicate dynamically stable (unstable) solutions. In panel (b), bistability occurs in the voltage window $V_1<V<V_2$. We take $\gamma/\Delta_0=0.1$, $T_{\rm env}=0$, and $\tau_{\rm imp}=\tau_{\rm mag}=\infty$ (clean limit). 
}
\label{Fig.BCS.Plead}
\end{figure}

Figure~\ref{Fig.BCS.Plead} shows steady-state solutions obtained by self-consistently solving the gap equation~\eqref{eq.noneq.gap} together with the steady-state condition~\eqref{eq.NESS}. Regarding this figure, we note that solving Eqs.~\eqref{eq.noneq.gap} and \eqref{eq.NESS} is not sufficient to determine the dynamical stability of a steady-state solution. In equilibrium superconductors, stability can be assessed from energetic considerations based on the free energy; however, this approach is not applicable to the present nonequilibrium case. Instead, the stability of a steady state must be examined by analyzing the time evolution of the order parameter~\cite{Snyman2009, Kawamura2024, Kawamura2025}. Such a dynamical stability analysis has been carried out in Ref.~\cite{Snyman2009}. From this previous work, one finds that while the solutions indicated by the solid lines in Fig.~\ref{Fig.BCS.Plead} are dynamically stable, those shown by the dashed lines are unstable.

In the case of symmetric coupling ($P_{\rm lead}=0$) shown in Fig.~\ref{Fig.BCS.Plead}(a), only a single stable NBCS solution is found to exist. With increasing the bias voltage, this superconducting solution disappears at $V_{\rm NBCS}^{\rm c}$, above which only the normal state with $\Delta=0$ remains. This critical voltage $V_{\rm NBCS}^{\rm c}$ corresponds to the blue dotted line in the phase diagram shown in Fig.~\ref{Fig.NSN}(c). Across this transition, the order parameter $\Delta$ drops discontinuously to zero, indicating a first-order–like phase transition from the superconducting state to the normal state.

\begin{figure}[t]
\centering
\includegraphics[width=\columnwidth]{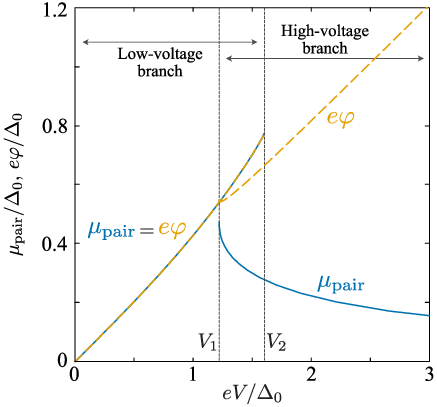}
\caption{
Calculated pair chemical potential $\mu_{\rm pair}$ (solid line) and electrostatic potential $\varphi$ (dashed line) in the asymmetric coupling case ($P_{\rm lead}=0.2$). The voltages $V_1$ and $V_2$ are the same as those in Fig.~\ref{Fig.BCS.Plead}(b).
}
\label{Fig.NBCS.mu.asy}
\end{figure}

When the coupling becomes asymmetric ($P_{\rm lead}\neq 0$), qualitatively new features emerge: As shown in Fig.~\ref{Fig.BCS.Plead}(b), an additional stable branch appears in the high-voltage regime ($V>V_1$), which is absent in the symmetric case. In particular, when $V_1<V<V_2$, two stable NBCS solutions coexist, that is, bistability occurs. Such bistable behavior has been reported previously~\cite{Snyman2009, Ouassou2018} and has also recently been observed experimentally~\cite{Shpagina2024, Shpagina2025}.

\begin{figure}[t]
\centering
\includegraphics[width=\columnwidth]{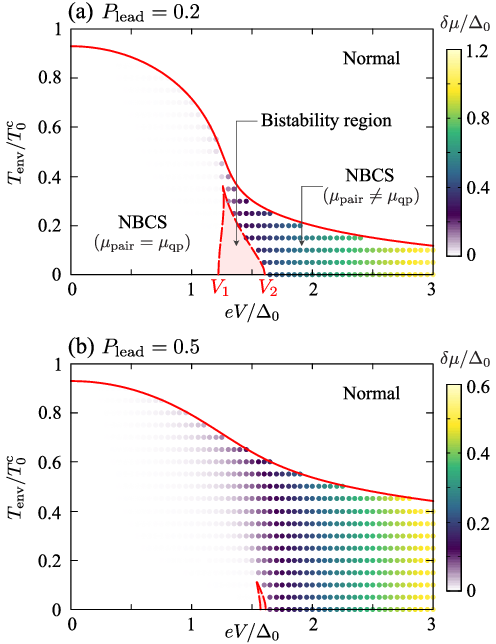}
\caption{
Phase diagram of the nonequilibrium superconductor for (a) $P_{\rm lead}=0.2$ and (b) $P_{\rm lead}=0.5$. The NBCS phase is color-coded according to the chemical potential difference $\delta\mu\equiv\mu_{\rm qp}-\mu_{\rm pair}$ between quasiparticles and the condensate. The red shaded area enclosed by the dashed lines ($V_1$ and $V_2$) is the bistability region.
}
\label{Fig.Phase}
\end{figure}

The two NBCS solutions in Fig.~\ref{Fig.BCS.Plead}(b) are distinguished by the presence or absence of the chemical-potential imbalance, $\mu_{\rm pair}\neq\mu_{\rm qp}$: Figure~\ref{Fig.NBCS.mu.asy} shows the voltage dependence of the pair chemical potential $\mu_{\rm pair}$ and the electrostatic potential $\varphi$ for $P_{\rm lead}=0.2$. For the low-voltage branch ($V<V_2$), one finds $\mu_{\rm pair}=e\varphi$. In this branch, since $\mu_{\rm qp}=e\varphi$, no chemical-potential imbalance exists between quasiparticles and the condensate. By contrast, the high-voltage branch ($V>V_1$) exhibits a clear imbalance between $\mu_{\rm pair}$ and $\mu_{\rm qp}$. We thus find that the NBCS state can be further classified into two types in the asymmetric coupling case ($P_{\rm lead}\neq 0$): one is accompanied by a chemical-potential imbalance, and the other is not. We emphasize that the latter-type NBCS state ($\mu_{\rm pair}\neq \mu_{\rm qp}$) is not obtained in the symmetric case, as seen in Fig.~\ref{Fig.BCS.Plead}(a).

This classification of the NBCS state is summarized in the phase diagram shown in Fig.~\ref{Fig.Phase}(a). In this phase diagram, the boundary between the normal and NBCS phases (solid line) coincides with the boundary obtained from the nonequilibrium Thouless criterion shown in Fig.~\ref{Fig.FFLO.asy}(d). As discussed in Sec.~\ref{sec.FFLO.asy}, the NFFLO state no longer exists when $P_{\rm lead}=0.2$, and the nonequilibrium superconducting state is dominated by the uniform NBCS state. The NBCS phase is color-coded according to the chemical potential difference $\mu_{\rm qp}-\mu_{\rm pair}$: The white region corresponds to $\mu_{\rm pair}=\mu_{\rm qp}$, and the colored region indicates a nonzero chemical-potential imbalance. In the red shaded area enclosed by dashed lines ($V_1<V<V_2$), the bistability phenomenon occurs. With increasing the temperature $T_{\rm env}$, this bistability region shrinks and eventually disappears when $T_{\rm env}/T^{\rm c}_0 \gtrsim 0.4$.

In the low-temperature regime ($T_{\rm env}/T^{\rm c}_0\lesssim0.4$), as the bias voltage is varied, a transition between the NBCS state with $\mu_{\rm pair}=\mu_{\rm qp}$ and that with $\mu_{\rm pair}\neq\mu_{\rm qp}$ occurs in the bistability region. The transition voltage depends on the sweep rate of the bias voltage~\cite{Snyman2009}. In the adiabatic limit, the transition occurs at $V=V_2$ ($V=V_1$) in the case where the voltage is increased (decreased). Since this transition is first-order–like, as shown in Fig.~\ref{Fig.BCS.Plead}(b), it is accompanied by a discontinuous change in the order-parameter amplitude $\Delta$.

We find from Fig.~\ref{Fig.Phase}(a) ($P_{\rm lead}=0.2$) and Fig.~\ref{Fig.Phase}(b) ($P_{\rm lead}=0.5$) that asymmetry between the two superconductor–lead couplings tends to suppress the bistability region. When the bias voltage is varied along a path that does not pass through the bistability region, the NBCS states with and without chemical-potential imbalance are no longer separated by a sharp transition. In this case, the superconducting order parameter varies smoothly without experiencing any discontinuity.

In summary, when $P_{\rm lead}\neq 0$, the NBCS state can be classified into two types, depending on whether the chemical-potential imbalance between quasiparticles and the condensate is present. At low temperatures, these two states are separated by a first–order–like transition accompanied by hysteresis, whereas at higher temperatures or in the strongly asymmetric case, they change smoothly from one to the other.


\subsection{Microscopic mechanism of the potential imbalance \label{sec.NBCS.imbalance}}

As shown in the preceding section, the NBCS state develops the chemical-potential imbalance ($\mu_{\rm pair}\neq \mu_{\rm qp}$) at high bias voltages in the asymmetric coupling case. In this subsection, we clarify its microscopic mechanism by revisiting the branch-imbalance picture, which was originally introduced by Tinkham~\cite{Tinkham1972_PRL, Tinkham1972}.

The chemical-potential imbalance $\mu_{\rm pair}\neq \mu_{\rm qp}$ in nonequilibrium superconductivity was first observed by Clarke~\cite{Clarke1972}. Subsequently, Tinkham developed a phenomenological theory and pointed out its essential origin as an imbalance between electron-like and hole-like quasiparticle populations~\cite{Tinkham1972_PRL, Tinkham1972, TinkhamBook}. In the following, we show that the imbalance observed in the NBCS state is naturally understood by this branch-imbalance picture. We also demonstrate that Tinkham’s phenomenological theory is fully consistent with our microscopic description based on the nonequilibrium Green’s function theory.

We note that a microscopic description of the chemical-potential imbalance has previously been formulated within the quasiclassical Green’s function technique~\cite{Schmid1975}. However, because quasiclassical theories integrate out the kinetic energy $\xi_{\bm{k}}$, their connection to Tinkham’s theory is not transparent. By contrast, our nonequilibrium Green’s function approach retains the explicit $\xi_{\bm{k}}$ dependence of physical quantities, which allows for a more direct and transparent discussion.

Within the BCS theory, the charge density $\rho$ in the thermal equilibrium superconducting state is given by
\begin{equation}
\rho
=
2e \sum_{\bm{k}}
\big[
u^2_{\bm{k}}\, f(E_{\bm{k}})
+
v^2_{\bm{k}}\, f(-E_{\bm{k}})
\big],
\label{eq.charge.BCS}
\end{equation}
where
\begin{equation}
u^2_{\bm{k}}
=
1-v^2_{\bm{k}}
=
\frac{1}{2}
\left[
1 + \frac{\xi_{\bm{k}}}{E_{\bm{k}}}
\right],
\end{equation}
$E_{\bm{k}}=\sqrt{\xi_{\bm{k}}^{2}+\Delta^2}$ is the Bogoliubov single-particle excitation energy, and $f(E_{\bm{k}})=[1+e^{E_{\bm{k}}/T}]^{-1}$ is the Fermi–Dirac distribution function. Equation~\eqref{eq.charge.BCS} can be rewritten as~\cite{Pethick1979, Pethick1980}
\begin{subequations}
\begin{align}
& \rho = \rho_{\rm qp} + \rho_{\rm pair},
\\[4pt]
& \rho_{\rm qp}
=
2\sum_{\bm{k}} q_{\bm{k}}\, f(E_{\bm{k}}),
\label{eq.rho.qp}
\\
& \rho_{\rm pair}
=
2e\sum_{\bm{k}} v^2_{\bm{k}}.
\end{align}
\label{eq.rho.CI}
\end{subequations}
Here, $\rho_{\rm qp}$ corresponds to the contribution from thermally excited quasiparticles, whereas $\rho_{\rm pair}$ arises from the coherent paired background~\cite{Leggett1977}. In Eq.~\eqref{eq.rho.qp}, we have introduced
\begin{equation}
q_{\bm{k}}
=
e\big[u^2_{\bm{k}} - v^2_{\bm{k}}\big]
=
e\, \frac{\xi_{\bm{k}}}{E_{\bm{k}}},
\end{equation}
which is known as the effective charge of quasiparticles~\cite{Pethick1979, Pethick1980}. Importantly, $q_{\bm{k}}$ is an odd function of the kinetic energy $\xi_{\bm{k}}$, while the distribution function $f(E_{\bm{k}})$ is an even function of $\xi_{\bm{k}}$. As a result, the momentum sum in Eq.~\eqref{eq.rho.qp} identically vanishes, yielding $\rho_{\rm qp}=0$. This result implies that, in the case of thermal-equilibrium superconductivity, the total electronic charge is entirely carried by the condensate at any temperature, and quasiparticles do \emph{not} contribute to the net charge density.

In the present nonequilibrium case, the counterpart of Eq.~\eqref{eq.rho.CI} can be derived from the nonequilibrium BCS theory, and the detailed derivation is presented in Appendix~\ref{sec.app.derivation.CI}. The resulting charge density in the nonequilibrium superconducting state consists of quasiparticle ($\rho_{\rm qp}$) and condensate ($\rho_{\rm pair}$) components, which are respectively given by
\begin{subequations}
\begin{align}
\rho_{\rm qp}
&=
2\sum_{\bm{k}} q_{\bm{k}}
\left[
f^{\rm s}_{\rm neq}(E_{\bm{k}}) -
\frac{\xi_{\bm{k}} +e\varphi}{E_{\bm{k}}}\,
f^{\rm a}_{\rm neq}(E_{\bm{k}})
\right],
\label{eq.rho.qp.noneq}
\\
\rho_{\rm pair}
&=
2e\sum_{\bm{k}} v^2_{\bm{k}},
\end{align}
\label{eq.rho.CI.noneq}
\end{subequations}
where
\begin{subequations}
\begin{align}
&
E_{\bm{k}} = \sqrt{[\xi_{\bm{k}} +e\varphi]^2 +\Delta^2}
\label{eq.Ek.phi}
,\\
&
u^2_{\bm{k}} = 1 -v^2_{\bm{k}} = 
\frac{1}{2}\left[1 +\frac{\xi_{\bm{k}} +e\varphi}{E_{\bm{k}}}\right]
\label{eq.ukvk.phi}
,\\
&
q_{\bm{k}} = e\, \big[u^2_{\bm{k}} -v^2_{\bm{k}}\big] =
e\, \frac{\xi_{\bm{k}} +e\varphi}{E_{\bm{k}}}
\label{eq.qk.phi}
,\\
&
f^{\rm a/s}_{\rm neq}(E_{\bm{k}}) = 
\frac{1}{2}\big[
f^+_{\rm neq}(E_{\bm{k}}) \pm
f^-_{\rm neq}(E_{\bm{k}}) \big]
\label{eq.def.fas}
,\\[4pt]
&
f^\pm_{\rm neq}(E_{\bm{k}}) = 
\frac{\gamma_1}{\gamma} f(E_{\bm{k}} \mp \mu_{\rm pair} \pm eV)
+
\frac{\gamma_2}{\gamma} f(E_{\bm{k}} \mp \mu_{\rm pair}).
\label{eq.def.fneq.pm}
\end{align}
\end{subequations}
Here, $E_{\bm{k}}$, $u^2_{\bm{k}}$, and $q_{\bm{k}}$ in Eqs.~\eqref{eq.Ek.phi}–\eqref{eq.qk.phi} have essentially the same form as in the thermal-equilibrium case, except for the replacement $\xi_{\bm{k}}\to\xi_{\bm{k}}+e\varphi$. This replacement accounts for the electrostatic potential shift in the nonequilibrium steady state. We also note that, in deriving Eq.~\eqref{eq.rho.CI.noneq}, we have taken the limit $\gamma\to+0$ while keeping the ratio $\gamma_1/\gamma_2$ fixed, for simplicity.

In Eq.~\eqref{eq.rho.qp.noneq}, the effective charge $q_{\bm{k}}$ is an odd function of $\xi_{\bm{k}}+e\varphi$, whereas $f^{\rm s}_{\rm neq}(E_{\bm{k}})$ is an even function of $\xi_{\bm{k}}+e\varphi$. As a result, the first term vanishes identically. In contrast, the second term, which originates from the nonequilibrium quasiparticle distribution and has no counterpart in the thermal-equilibrium case, generally yields a nonzero contribution, resulting in a nonzero quasiparticle charge density $\rho_{\rm qp}\neq 0$.

\begin{figure}[t]
\centering
\includegraphics[width=\columnwidth]{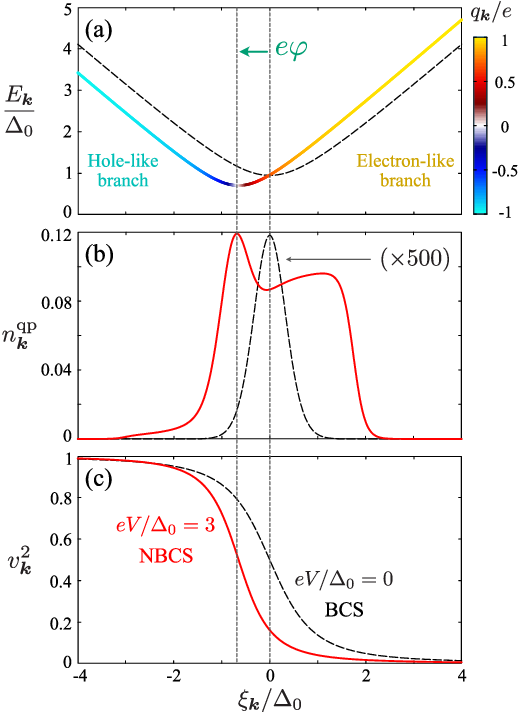}
\caption{
Comparison of the thermal-equilibrium BCS state ($eV/\Delta_0=0$) with the NBCS state ($eV/\Delta_0=3$). We set $P_{\rm lead}=0.8$ and $T_{\rm env}/T_0^{\rm c}=0.2$. Results for the NBCS (BCS) state are shown by solid (dashed) lines. (a) Quasiparticle excitation spectrum $E_{\bm{k}}$. The NBCS spectrum is color-coded according to the quasiparticle effective charge $q_{\bm{k}}$. (b) Quasiparticle momentum distribution $n_{\bm{k}}^{\rm qp}$. For clarity, the BCS result is multiplied by a factor of 500. (c) Condensate distribution $v_{\bm{k}}^{2}$.
}
\label{Fig.CI}
\end{figure}

\begin{figure}[t]
\centering
\includegraphics[width=\columnwidth]{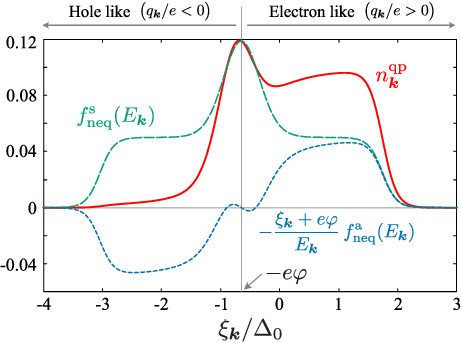}
\caption{
Decomposition of the NBCS quasiparticle momentum distribution $n_{\bm{k}}^{\rm qp}$ shown in Fig.~\ref{Fig.CI}(b). The dashed and dotted lines represent the contributions from the first and second terms in Eq.~\eqref{eq.nqp.noneq}, respectively.
}
\label{Fig.CI.nqp}
\end{figure}

To illustrate this mechanism more explicitly, Fig.~\ref{Fig.CI} compares the thermal-equilibrium BCS state ($eV/\Delta_0=0$) with the NBCS state ($eV/\Delta_0=3$), when $P_{\rm lead}=0.8$ and $T_{\rm env}/T_0^{\rm c}=0.2$. Figure~\ref{Fig.CI}(a) shows the quasiparticle excitation spectrum $E_{\bm{k}}$. In the NBCS state, a nonzero electrostatic potential $\varphi$ shifts the excitation spectrum by $-e\varphi$ relative to that in the thermal-equilibrium state (dashed line). The NBCS spectrum is color-coded by the quasiparticle effective charge $q_{\bm{k}}$ defined in Eq.~\eqref{eq.qk.phi}. Each momentum state carries a different electric charge: quasiparticles with $\xi_{\bm{k}}+e\varphi>0$ carry charge with the same sign as electrons ($q_{\bm{k}}/e>0$), whereas those with $\xi_{\bm{k}}+e\varphi<0$ carry charge with the same sign as holes ($q_{\bm{k}}/e<0$). Thus, these parts of the spectrum are referred to as the electron-like and hole-like branches, respectively~\cite{Tinkham1972_PRL, Tinkham1972, TinkhamBook}.

Figure~\ref{Fig.CI}(b) shows the quasiparticle distribution $n_{\bm{k}}^{\rm qp}$, which describes the occupation of the excitation spectrum $E_{\bm{k}}$. As seen from Eqs.~\eqref{eq.rho.qp} and~\eqref{eq.rho.qp.noneq}, one has $n_{\bm{k}}^{\rm qp}=f(E_{\bm{k}})$ in the thermal-equilibrium state, whereas the NBCS state gives
\begin{equation}
n_{\bm{k}}^{\rm qp}=
f^{\rm s}_{\rm neq}(E_{\bm{k}}) -
\frac{\xi_{\bm{k}} +e\varphi}{E_{\bm{k}}}\,
f^{\rm a}_{\rm neq}(E_{\bm{k}}).
\label{eq.nqp.noneq}
\end{equation}
In the thermal-equilibrium case (dashed line), the quasiparticle distribution is symmetric with respect to $\xi_{\bm{k}}=0$, so that the electron-like ($\xi_{\bm{k}}>0$) and hole-like ($\xi_{\bm{k}}<0$) branches are equally populated. As a result, the net quasiparticle contribution to the charge density vanishes. By contrast, in the NBCS state, the distribution becomes asymmetric, so that the electron-like branch ($\xi_{\bm{k}}+e\varphi>0$) is more strongly populated than the hole-like branch ($\xi_{\bm{k}}+e\varphi<0$). This population imbalance gives rise to a nonzero quasiparticle charge density.

The origin of the branch imbalance becomes transparent by decomposing the quasiparticle distribution $n_{\bm{k}}^{\rm qp}$ in Eq.~\eqref{eq.nqp.noneq} into its first and second terms. Figure~\ref{Fig.CI.nqp} shows this decomposition of the total distribution $n_{\bm{k}}^{\rm qp}$ presented in Fig.~\ref{Fig.CI}(b). The first term (dashed line) excites electron-like and hole-like quasiparticles equally and therefore does not contribute to the net charge density. By contrast, the second term (dotted line) induces a branch imbalance, with a higher population of electron-like quasiparticles than hole-like ones.

When quasiparticles carry electric charge, charge neutrality of the system, $\rho=0$, requires compensation of this excess charge by the condensate $\rho_{\rm pair}$. As shown in Fig.~\ref{Fig.CI}(c), the condensate distribution $v_{\bm{k}}^{2}$ in the NBCS state is shifted by $-e\varphi$ relative to that in the thermal-equilibrium state. The area between the solid and dashed curves represents the reduction of $\rho_{\rm pair}$, which exactly compensates the excess charge carried by quasiparticles.

To summarize, the nonequilibrium quasiparticle distribution induced by quasiparticle injection generates a branch imbalance and endows quasiparticles with a nonzero electric charge. To satisfy the charge neutrality of the system, the condensate distribution $v_{\bm{k}}^{2}$ adjusts accordingly, leading to a reduction of $\rho_{\rm pair}$ and the emergence of a nonzero electrostatic potential $\varphi\neq 0$. Since the chemical potential of Cooper pairs is chosen as the energy reference ($\mu_{\rm pair}=0$) in the nonequilibrium BCS theory, the condition $\mu_{\rm qp}=e\varphi\neq 0$ immediately means a chemical-potential imbalance between quasiparticles and the condensate. This provides the microscopic origin of the chemical-potential difference $\mu_{\rm pair}\neq \mu_{\rm qp}$ in the NBCS state at high bias voltages.

We note that the symmetric coupling ($\gamma_1=\gamma_2$) enforces $\mu_{\rm pair}=eV/2$. In this case, $f^{+}_{\rm neq}(E_{\bm{k}})=f^{-}_{\rm neq}(E_{\bm{k}})$ holds, giving $f^{\rm a}_{\rm neq}(E_{\bm{k}})=0$. As a result, the branch-imbalanced contribution vanishes, and no chemical-potential (charge) imbalance is generated. This conclusion is consistent with previous work indicating that charge imbalance does not occur in a symmetrically coupled NSN tunneling junction~\cite{Takahashi1999}.

Finally, we comment on the relation between Eq.~\eqref{eq.rho.qp.noneq} and Tinkham’s expression~\cite{Tinkham1972_PRL, Tinkham1972}. In Tinkham’s treatment, the quasiparticle contribution to the charge density ($eQ^{*}$ in his notation) is written as
\begin{equation}
\rho_{\rm qp}
=
2eN(0)\int_{\Delta}^{\infty} dE\,
\big[
\delta f_{k>}(E)-\delta f_{k<}(E)
\big],
\label{eq.Tinkham}
\end{equation}
where $\delta f_{k>}$ and $\delta f_{k<}$ denote nonequilibrium distribution functions introduced for the electron-like ($k>k_{\rm F}$) and hole-like ($k<k_{\rm F}$) branches, respectively. In our scheme, the first term in Eq.~\eqref{eq.rho.qp.noneq} vanishes, and the remaining contribution can be expressed as
\begin{align}
\rho_{\rm qp}
=
2eN(0)\int_0^\infty d\xi'\,
\left(\frac{\xi'}{E^\prime}\right)^2
\big[
f^{-}_{\rm neq}(E')-
f^{+}_{\rm neq}(E')
\big],
\label{eq.rhoqp.xi.prime}
\end{align}
where $\xi'=\xi_{\bm{k}}+e\varphi$ and $E'=\sqrt{\xi'^2+\Delta^2}$. Changing the integration variable from $\xi'$ to $E'$ in Eq.~\eqref{eq.rhoqp.xi.prime}, one reaches
\begin{equation}
\rho_{\rm qp}
=
2eN(0)\int_{\Delta}^{\infty} dE\,
\sqrt{1-\left(\frac{\Delta}{E}\right)^2}
\big[
f^{-}_{\rm neq}(E)-f^{+}_{\rm neq}(E)
\big].
\end{equation}
Comparing this expression with Eq.~\eqref{eq.Tinkham}, we identify the correspondence
\begin{equation}
\delta f_{k\gtrless}(E)
=
\sqrt{1-\left(\frac{\Delta}{E}\right)^2}\,
f^{\mp}_{\rm neq}(E).
\label{eq.fneq.fgtrless}
\end{equation}
Equation~\eqref{eq.fneq.fgtrless} clearly shows that Tinkham’s phenomenological description of potential (charge) imbalance is consistent with our microscopic formulation based on the nonequilibrium Green’s function technique.

\section{Effects of Impurity Scattering on Nonequilibrium Superconductivity \label{sec.imp*}}

\subsection{Suppression of the NFFLO State by Impurity Scattering \label{sec.imp}}

As discussed in Sec.~\ref{sec.FFLO.asy}, asymmetry between the two superconductor–lead couplings is detrimental to the NFFLO state. In addition, impurity scattering also plays a crucial role in determining its stability.

\begin{figure}[t]
\centering
\includegraphics[width=\columnwidth]{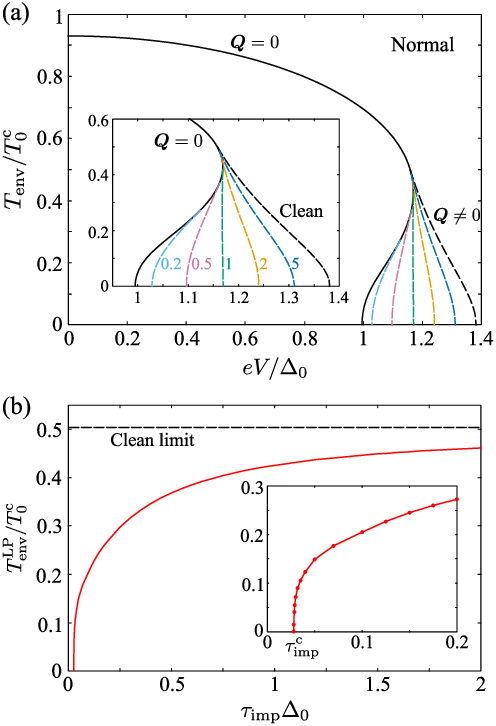}
\caption{
(a) Superconducting phase transition temperature $T_{\rm env}^{\rm c}$ as a function of bias voltage $V$ in the symmetric case ($P_{\rm lead}=0$ and $\gamma/\Delta_0=0.1$). The solid and dashed lines represent the NBCS ($\bm{Q}=0$) and NFFLO ($\bm{Q}\neq 0$) transitions, respectively. Each dashed line shows the NFFLO phase transition temperature in the presence of nonmagnetic impurities ($0.2 \leq \tau_{\rm imp}\Delta_0 \leq 5$). We note that the NBCS phase transition temperature (solid line) is independent of $\tau_{\rm imp}$, as will be discussed in Sec.~\ref{sec.imp.NBCS}. The inset shows the same figure for $eV/\Delta_0=[0.95, 1.4]$. (b) Lifshitz-point temperature $T_{\rm env}^{\rm LP}$ as a function of the relaxation time $\tau_{\rm imp}$. $T_{\rm env}^{\rm LP}$ vanishes at $\tau_{\rm imp}^{\rm c}\Delta_0\simeq 0.028$. The inset shows the same figure for $\tau_{\rm imp}\Delta_0 = [0, 0.2]$. 
}
\label{Fig.FFLO.imp}
\end{figure}

Figure~\ref{Fig.FFLO.imp}(a) shows the calculated superconducting phase transition temperature $T_{\rm env}^{\rm c}$ in the symmetric case ($P_{\rm lead}=0$). While the solid line represents the transition from the normal state to the NBCS state ($\bm{Q}=0$), the dashed line shows the transition to the NFFLO state ($\bm{Q}\neq 0$).  These two lines correspond to the red solid and green dashed lines in Fig.~\ref{Fig.NSN}(c), respectively.

As will be discussed in Sec.~\ref{sec.imp.NBCS}, the NBCS transition temperature is insensitive to nonmagnetic impurity scattering. Thus, the solid line in Fig.~\ref{Fig.FFLO.imp}(a) is independent of the relaxation time $\tau_{\rm imp}$. In contrast, the dashed lines in Fig.~\ref{Fig.FFLO.imp}(a) show that the NFFLO state is strongly suppressed by impurity scattering.

This tendency becomes more transparent when we focus on the Lifshitz-point temperature $T_{\rm env}^{\rm LP}$, at which the leading instability changes from the uniform NBCS type ($\bm{Q}=0$) to the NFFLO type ($\bm{Q}\neq 0$). As seen in Fig.~\ref{Fig.FFLO.imp}(b), $T_{\rm env}^{\rm LP}$ is suppressed with decreasing $\tau_{\rm imp}$ and eventually vanishes when $\tau_{\rm imp}\Delta_0 \lesssim 0.028$. This means that the NFFLO instability no longer occurs when $\tau_{\rm imp} < \tau_{\rm imp}^{\rm c}$.

The origin of this suppression of the NFFLO phase transition by nonmagnetic impurities is essentially the same as that of the FFLO state in the thermal-equilibrium case under an external magnetic field~\cite{Matsuda2007}. In the NFFLO state, Cooper pairs are formed between electrons around different effective Fermi surfaces with momenta $(\bm{k}+\bm{Q},\sigma)$ and $(-\bm{k},-\sigma)$, as schematically illustrated in Fig.~\ref{Fig.NSN}(b). When these electrons are scattered by impurities to $(\bm{k}'+\bm{Q},\sigma)$ and $(-\bm{k}',-\sigma)$, they no longer remain simultaneously around the effective Fermi surfaces, which naturally acts as a depairing effect on Cooper pairs. Thus, as in the case of the thermal-equilibrium FFLO state, the realization of the NFFLO state also requires sufficiently clean samples.

\subsection{Robustness of the NBCS State against Nonmagnetic Impurity Scattering \label{sec.imp.NBCS}}

In this section, we show that nonmagnetic impurity scattering does not suppress the NBCS state, in contrast to the NFFLO case (Sec.~\ref{sec.imp}). In the thermal-equilibrium case, the robustness of an $s$-wave BCS superconductor against nonmagnetic impurities is well known as Anderson's theorem~\cite{Anderson1959}. Here, we demonstrate that an analogous robustness persists even in the present nonequilibrium situation.

We first examine the critical temperature $T_{\rm env}^{\rm c}$ of the NBCS state. The critical temperature $T_{\rm env}^{\rm c}$ is determined by solving the Thouless criterion in Eqs.~\eqref{eq.Thouless.Re} and \eqref{eq.Thouless.Im} for the uniform pairing channel $\bm{Q}=0$. Using the identity
\begin{equation}
\lim_{q\to +0}\, \frac{1}{2\bar{q}} 
\log\left(\frac{\omega+\nu/2 \pm i\tilde{\gamma}/2 +\bar{q}}{\omega+\nu/2 \pm i\tilde{\gamma}/2 -\bar{q}}
\right)
=
\frac{1}{\omega+\nu/2 \pm i\tilde{\gamma}/2},
\end{equation}
one finds that the pair correlation function in Eq.~\eqref{eq.PiR}, as well as the vertex function in Eq.~\eqref{eq.vertex}, reduce to, respectively,
\begin{align}
&
\Pi^{\rm R}(\bm{0},\nu)=
\frac{N(0)}{4}\int_{-\infty}^\infty d\omega\,
\notag\\
&\hspace{2cm}
\times
\Bigg[
\frac{1 -2f_{\rm neq}(-\omega)}{\omega +\nu/2 +i\tilde{\gamma}/2}\,
\Lambda(\bm{0},\omega +\nu, -\omega)
\notag\\
&\hspace{2cm}
-
\frac{1 -2f_{\rm neq}(\omega+\nu)}{\omega +\nu/2 -i\tilde{\gamma}/2}\, \Lambda^*(\bm{0},\omega +\nu, -\omega)
\Bigg]
\label{eq.piR.q0}
,\\
&
\Lambda(\bm{0},\omega +\nu, -\omega)=
\frac{\omega+\nu/2 +i\tilde{\gamma}}{\omega+\nu/2+i\gamma}.
\label{eq.vertex.q0}
\end{align}
Substituting Eq.~\eqref{eq.vertex.q0} into Eq.~\eqref{eq.piR.q0}, one has
\begin{align}
&
\Pi^{\rm R}(\bm{0},\nu)=
\frac{N(0)}{4}\int_{-\infty}^\infty d\omega
\notag\\
&\hspace{1cm}
\times
\Bigg[
\frac{1 -2f_{\rm neq}(-\omega)}{\omega +\nu/2 +i\gamma/2}
-\frac{1 -2f_{\rm neq}(\omega+\nu)}{\omega +\nu/2 -i\gamma/2}
\Bigg].
\label{eq.PiR.q0.2}	
\end{align}
Remarkably, the impurity-renormalized scattering rate $\tilde{\gamma}$ defined in Eq.~\eqref{eq.til.gamma} drops out, and the Thouless criterion in Eqs.~\eqref{eq.Thouless.Re} and \eqref{eq.Thouless.Im} for the uniform pairing channel ($\bm{Q}=0$) is entirely insensitive to the nonmagnetic impurity scattering rate $\tau_{\rm imp}$. This implies that the critical temperature $T_{\rm env}^{\rm c}$ of the NBCS state remains unaffected by nonmagnetic impurity scattering.

This vanishing nonmagnetic impurity effect persists over the entire NBCS phase. To demonstrate this within the nonequilibrium BCS theory, we rewrite the Dyson equation~\eqref{eq.Dyson.Nambu} as
\begin{subequations}
\begin{align}
&
\hat{\bm{G}}_{\rm clean}(\bm{k},\omega)
=
\hat{\bm{G}}_{0}(\bm{k},\omega)
\notag\\
&\hspace{0.3cm}
+
\hat{\bm{G}}_{0}(\bm{k},\omega)\,
\big[
\hat{\bm{\Sigma}}_{\rm int}(\omega)+
\hat{\bm{\Sigma}}_{\rm lead}(\omega)\big]\,
\hat{\bm{G}}_{\rm clean}(\bm{k},\omega),
\label{eq.Dyson_clean}
\\[6pt]
&
\hat{\bm{G}}(\bm{k},\omega)
=
\hat{\bm{G}}_{\rm clean}(\bm{k},\omega)
+
\hat{\bm{G}}_{\rm clean}(\bm{k},\omega)\,
\hat{\bm{\Sigma}}_{\rm imp}(\omega)\,
\hat{\bm{G}}(\bm{k},\omega).
\label{eq.Dyson_imp}
\end{align}
\label{eq.Dyson.clean.imp}
\end{subequations}
Here, we ignore the effects of magnetic impurities, $\hat{\bm{\Sigma}}_{\rm mag}$. In Eq.~\eqref{eq.Dyson.clean.imp}, $\hat{\bm{G}}_{\rm clean}$ denotes the ``clean-limit'' Green’s function, which incorporates the effects of the pairing interaction as well as the coupling to the leads, but excludes the effects of nonmagnetic impurities.

The Keldysh component of $\hat{\bm{G}}_{\rm clean}$ can generally be written as~\cite{Schmid1975, Rammer2007}
\begin{equation}
\bm{G}^{\rm K}_{\rm clean}(\bm{k},\omega)
=
\bm{G}^{\rm R}_{\rm clean}(\bm{k},\omega)\,\bm{h}(\omega)
-
\bm{h}(\omega)\,
\bm{G}^{\rm A}_{\rm clean}(\bm{k},\omega),
\label{eq.GK_clean_h}
\end{equation}
where $\bm{h}(\omega)$ represents the nonequilibrium electron distribution, which can be conveniently decomposed into the longitudinal (L) and transverse (T) modes as~\cite{Schmid1975, Rammer2007}
\begin{equation}
\bm{h}(\omega) = 
f_{\rm L}(\omega)\, \big(\tau_0 \otimes \sigma_0 \big)
+
f_{\rm T}(\omega)\, \big(\tau_3 \otimes \sigma_0 \big).
\end{equation}
As shown in Appendix~\ref{sec.app.proof}, the impurity-dressed Keldysh Green’s function also obeys the same equation as Eq.~\eqref{eq.GK_clean_h}, that is,
\begin{equation}
\bm{G}^{\rm K}(\bm{k},\omega)
=
\bm{G}^{\rm R}(\bm{k},\omega)\,
\bm{h}(\omega)
-
\bm{h}(\omega)\,
\bm{G}^{\rm A}(\bm{k},\omega),
\label{eq.GK_full_h}
\end{equation}
with the \emph{identical} distribution function $\bm{h}(\omega)$ appearing in Eq.~\eqref{eq.GK_clean_h}. Physically, this reflects the fact that elastic scattering from impurities does not induce energy relaxation, and therefore leaves the nonequilibrium electron distribution unchanged~\cite{Kawamura2026}.

We now explicitly show that the nonequilibrium gap equation~\eqref{eq.noneq.gap} is unchanged by nonmagnetic impurity scattering: Substituting Eq.~\eqref{eq.GK_clean_h} into Eq.~\eqref{eq.noneq.gap}, we obtain the clean-limit gap equation,
\begin{align}
\Delta
&=
-\frac{i U N(0)}{2}\int_{-\omega_{\rm D}}^{\omega_{\rm D}}
\frac{d\omega}{2\pi}\, f_{\rm L}(\omega)
\notag\\
&\hspace{1cm}\times
{\rm Tr}\Big[
\big(\tau_- \otimes \sigma_-\big)\,
\big[
\bm{g}^{\rm R}_{\rm clean}(\omega)
-
\bm{g}^{\rm A}_{\rm clean}(\omega)
\big]
\Big]
\notag\\[6pt]
&
+\frac{i U N(0)}{2}\int_{-\omega_{\rm D}}^{\omega_{\rm D}}
\frac{d\omega}{2\pi}\, f_{\rm T}(\omega)
\notag\\
&\hspace{1cm}\times
{\rm Tr}\Big[
\big(\tau_- \otimes \sigma_-\big)\,
\big[
\bm{g}^{\rm R}_{\rm clean}(\omega)
+
\bm{g}^{\rm A}_{\rm clean}(\omega)
\big]
\Big],
\label{eq.gap_clean}
\end{align}
where
\begin{equation}
\bm{g}^{\rm R}_{\rm clean}(\omega)
=
\big[\bm{g}^{\rm A}_{\rm clean}(\omega)\big]^\dagger
=
-\pi\,
\frac{\omega_+ + \Delta\,(\tau_2 \otimes \sigma_y)}
{\sqrt{\Delta^2-\omega_+^2}}.
\label{eq.gr_clean}
\end{equation}
As shown in Eq~\eqref{eq.GK_full_h}, the dressed Keldysh Green’s function $\bm{G}^{\rm K}$ obeys the same equation as that in the clean limit in Eq.~\eqref{eq.GK_clean_h}. Thus, in the presence of nonmagnetic impurities, the gap equation is obtained from the clean-limit expression in Eq.~\eqref{eq.gap_clean} by simply replacing the clean-limit Green’s functions $\bm{g}^{\rm R(A)}_{\rm clean}$ with the impurity-dressed ones $\bm{g}^{\rm R(A)}$ given in Eq.~\eqref{eq.gr.qc}.

The relation between $\bm{g}^{\rm R(A)}_{\rm clean}$ and $\bm{g}^{\rm R(A)}$ is obtained from Eq.~\eqref{eq.til}. In the absence of magnetic impurities ($\tau_{\rm mag}\to\infty$), Eq.~\eqref{eq.til} implies that the impurity-renormalized frequency $\tilde{\omega}_+$ and the gap function $\tilde{\Delta}(\omega)$ satisfy the relation,
\begin{equation}
\frac{\tilde{\omega}_+}{\omega_+}
=
\frac{\tilde{\Delta}(\omega)}{\Delta}.
\label{eq.til_relation}
\end{equation}
One then readily finds
\begin{equation}
\bm{g}^{\rm R(A)}(\omega)
=
\bm{g}^{\rm R(A)}_{\rm clean}(\omega).
\label{eq.gR_gA_identity}
\end{equation}
As a consequence, all impurity-induced renormalizations drop out, and the nonequilibrium gap equation~\eqref{eq.noneq.gap} remains identical to the clean-limit case, even in the presence of nonmagnetic impurities. The same argument is applicable to the steady-state condition in Eq.~\eqref{eq.NESS}. Thus, neither the superconducting gap $\Delta$ nor the pair chemical potential $\mu_{\rm pair}$ in the NBCS state is affected by nonmagnetic impurity scattering. It follows that the results in Sec.~\ref{sec.NBCS.steady} remain unchanged even in the presence of nonmagnetic impurities ($\tau_{\rm imp}^{-1}\neq 0$).

The above discussion indicates that the robustness of the NBCS state against nonmagnetic impurities arises because they do not affect the superconducting density of states,
\begin{equation}
\frac{N_{\rm S}(\omega)}{N(0)} = -\frac{1}{2\pi} {\rm Im}\, {\rm Tr}\Big[\big(\tau_{\rm p}\otimes \sigma_0\big)\, \bm{g}^{\rm R}(\omega)\Big],
\end{equation}
as well as the nonequilibrium distribution function $\bm{h}(\omega)$. This also implies that perturbations that alter either of these quantities would affect the NBCS state. Indeed, as will be shown in Sec.~\ref{sec.NBCS.magimp}, magnetic impurity scattering, which alters the superconducting density of states $N_{\rm S}(\omega)$, strongly suppresses the NBCS state. Furthermore, although not considered in the present work, inelastic scattering processes, such as electron–phonon scattering, are expected to influence the NBCS state, since they induce energy redistribution and thereby alter the nonequilibrium distribution function $\bm{h}(\omega)$~\cite{Kawamura2026}.

\subsection{Effects of Magnetic Impurity Scattering on the NBCS State
 \label{sec.NBCS.magimp}}

\begin{figure}[t]
\centering
\includegraphics[width=\columnwidth]{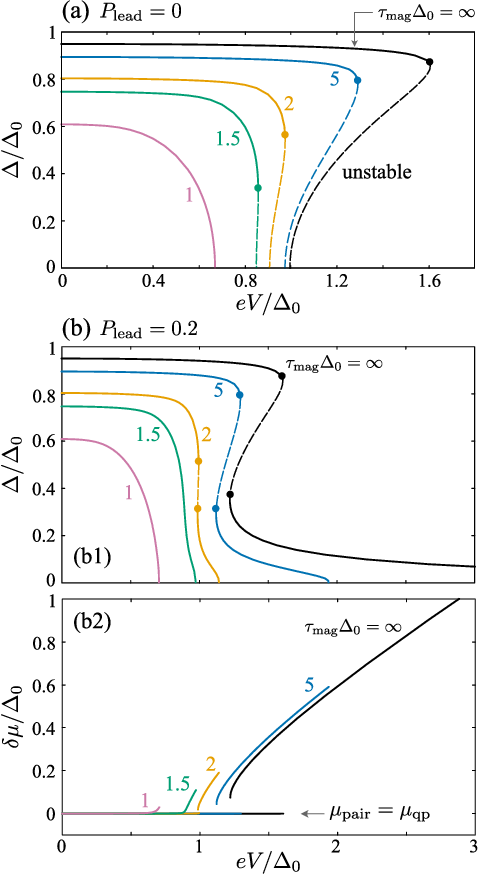}
\caption{
Calculated superconducting order parameter $\Delta$ of the NBCS state in the presence of magnetic impurities. We take $\gamma/\Delta_0=0.1$ and $T_{\rm env}=0$. (a) Symmetric lead-coupling case ($P_{\rm lead}=0$). (b1) Asymmetric lead-coupling case ($P_{\rm lead}=0.2$). Solid and dashed lines indicate dynamically stable and unstable solutions, respectively, and the solid circle on each line marks the boundary between these two states. (b2) Chemical potential difference $\delta\mu=\mu_{\rm qp}-\mu_{\rm pair}$ of the stable NBCS state in the asymmetric lead-coupling case.
}
\label{Fig.BCS.mag}
\end{figure}

Figure~\ref{Fig.BCS.mag} shows how magnetic impurity scattering affects the superconducting order parameter $\Delta$ of the NBCS state. (Note that the result in the limit $\tau_{\rm mag}\to\infty$ coincides with that shown in Fig.~\ref{Fig.BCS.Plead}.) When $P_{\rm lead}=0$ (symmetric case), Fig.~\ref{Fig.BCS.mag}(a) shows that the NBCS state is suppressed by magnetic impurity scattering, as in the case of the thermal-equilibrium BCS state~\cite{Maki1969}. We also find from this figure that the discontinuous phase transition to the normal state at $V=V_{\rm NBCS}^{\rm c}$ in the absence of magnetic impurities disappears when $\tau_{\rm mag}\Delta_0 = 1$, where the transition to the normal state becomes continuous.

In the asymmetric coupling case ($P_{\rm lead}=0.2$) shown in Fig.~\ref{Fig.BCS.mag}(b1), besides the suppression of the superconducting order parameter $\Delta$, the presence of magnetic impurities is also found to suppress the bistability phenomenon seen in Fig.~\ref{Fig.BCS.Plead}(b). We also find from Fig.~\ref{Fig.BCS.mag}(b2) that the chemical potential difference $\delta \mu=\mu_{\rm qp}-\mu_{\rm pair}$ in the charge-imbalanced NBCS state is strongly suppressed by magnetic impurities. Thus, magnetic impurities are unfavorable for observing the superconducting phase transition, the bistability phenomenon, as well as the charge-imbalance phenomenon of the NBCS state.

The qualitative difference between the effects of nonmagnetic and magnetic impurities originates from their distinct influences on the superconducting density of states $N_{\rm S}(\omega)$. As explained in Sec.~\ref{sec.imp.NBCS}, while nonmagnetic impurities do not modify $N_{\rm S}(\omega)$, magnetic impurities directly affect $N_{\rm S}(\omega)$ because the relation given by Eq.~\eqref{eq.til_relation} no longer holds. This modification in the latter case suppresses the NBCS state and alters the nonequilibrium phase-transition behavior.

\section{Summary}

In this work, we have investigated nonequilibrium superconductivity in a voltage-biased normal metal–superconductor–normal metal (NSN) junction, with particular emphasis on the effects of asymmetry between the two superconductor–lead couplings and impurity scattering. Using the Keldysh Green’s function technique, we extended the thermal-equilibrium mean-field BCS theory to the nonequilibrium case. Using this approach, we clarified the nonequilibrium superconducting phase diagram beyond previous work~\cite{Kawamura2024}, which was restricted to idealized conditions with symmetric lead coupling and the clean limit.

We found that the inhomogeneous nonequilibrium Fulde--Ferrell--Larkin--Ovchinnikov (NFFLO) state induced by the nonequilibrium electron distribution is fragile against impurity scattering, in close analogy with the equilibrium FFLO state. We also found that asymmetry between the two superconductor–lead couplings is detrimental to the stability of the NFFLO state. These results indicate that the realization of the NFFLO state requires clean samples and nearly symmetric coupling to the two normal-metal leads.

In the asymmetric lead-coupling case, we showed that the uniform nonequilibrium BCS (NBCS) state can be classified into two types: one possesses a nonzero chemical-potential imbalance between quasiparticles and the condensate, and the other does not have such an imbalance. We found a phase transition or crossover between these two NBCS states and determined the region in the phase diagram where they exhibit bistability. Moreover, we demonstrated that the emergence of the chemical-potential imbalance can be naturally understood in terms of a population imbalance between electron-like and hole-like quasiparticles induced by the nonequilibrium electron distribution.

\begin{figure}[t]
\centering
\includegraphics[width=\columnwidth]{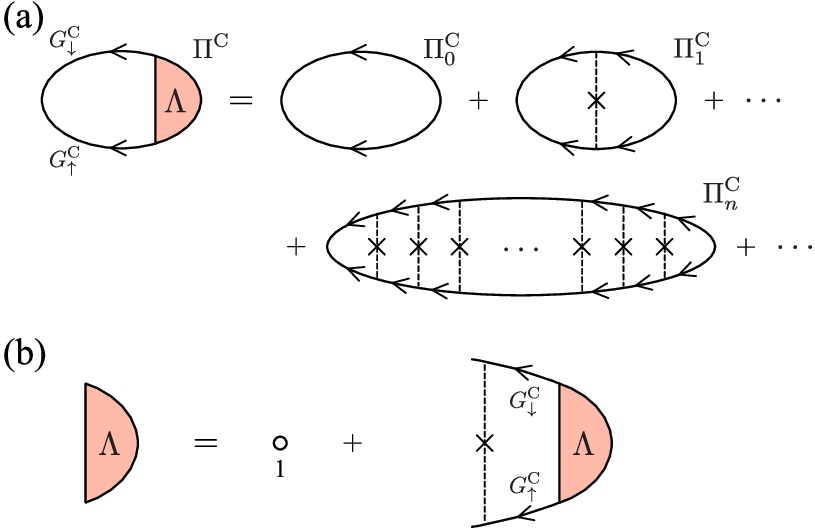}
\caption{
Diagrammatic representation of (a) the contour-ordered pair correlation function $\Pi^{\rm C}$ and (b) the equation for the three-point vertex function $\Lambda$. The solid line represents the dressed contour-ordered Green’s function $G^{\rm C}_\sigma$, including self-energy corrections due to impurity scattering as well as coupling to the normal-metal leads. The dotted line with a cross represents impurity scattering.
}
\label{Fig.diagram}
\end{figure}

Regarding the stability of these uniform NBCS states, we further showed that they are robust against nonmagnetic impurity scattering, as in the case of the thermal-equilibrium BCS state, whereas they are suppressed by magnetic impurities. Thus, magnetic impurities should be minimized in order to observe the phase transition and bistability between the NBCS states with and without the chemical-potential imbalance.

Our results provide a comprehensive understanding of nonequilibrium superconducting properties in NSN junctions under realistic conditions, and are expected to stimulate further studies of nonequilibrium phenomena in a broad class of superconducting junctions. As one example, ferromagnet–superconductor–ferromagnet (FSF) junctions have recently attracted much attention in the context of superconducting spintronics~\cite{Takahashi1999, Takahashi2002, Linder2015, Zhao1995, Hubler2012, Quay2013, Silaev2015, Kolenda2016, Maier2023}. In FSF systems, the interplay between potential (charge) imbalance and spin imbalance is actively discussed~\cite{Zhao1995, Takane2006, Hubler2012, Quay2013, Silaev2015, Kolenda2016, Maier2023}. We expect that the theoretical framework developed in this work will also be useful for addressing such problems.

\begin{acknowledgments}
We thank Shuntaro Sumita, Yusuke Kato, and H.T.C. Stoof for stimulating discussions. We are particularly grateful to Yusuke Kato for his insightful and informative comments on charge imbalance in nonequilibrium superconductivity. T.K. was supported by MEXT and JSPS KAKENHI Grant-in-Aid for JSPS fellows Grant No.~25K23363. Y.O. was supported by No.~JP22K03486. Some of the computations in this work were done using the facilities of the Supercomputer Center, the Institute for Solid State Physics, the University of Tokyo.
\end{acknowledgments}

\appendix
\section{Derivation of Eqs.~\eqref{eq.PiR} and \eqref{eq.BSeq} \label{sec.app.Pi}}

We define the contour-ordered pair correlation function, which is the lowest-order contribution with respect to the attractive interaction $-U$, as
\begin{equation}
\Pi^{\rm C}(\bm{q},z, z')= -i \braket{\hat{T}_{\rm C}\, \m{B}_{\bm{q}}(z)\, \m{B}^\dagger_{-\bm{q}}(z')}.
\end{equation}
Here, $z$ denotes contour-time variables on the Keldysh contour ${\rm C}$ and $\hat{T}_{\rm C}$ is the contour-order operator~\cite{Rammer2007, Stefanucci2013, Haug2008}. The pair annihilation operator is defined by
\begin{equation}
\m{B}_{\bm{q}} = \sum_{\bm{k}} 
a_{-\bm{k}+\bm{q}/2, \down}a_{\bm{k}+\bm{q}/2, \up}.
\end{equation}
We perturbatively evaluate $\Pi^{\rm C}$ in terms of the impurity-scattering Hamiltonian $H_{\rm imp}$ in Eq.~\eqref{eq.H.imp}, as well as the tunneling Hamiltonian $H_{\rm mix}$ in Eq.~\eqref{eq.H.lead}. Since $\Pi^{\rm C}$ is a two-particle Green's function, impurity effects generally enter not only through the self-energy but also through vertex corrections.

In this paper, impurity scattering effects on the single-particle Green’s function are treated within the self-consistent Born approximation. In this case, consistency with the Ward–Takahashi identity~\cite{Ward1950, Takahashi1957} requires the inclusion of the impurity ladder-type vertex corrections~\cite{Mahan2000}, which is diagrammatically given as Fig.~\ref{Fig.diagram}. Evaluating these diagrams and taking spatial average over the impurity positions $\bm{R}_j^{\rm i}$~\cite{Rammer2007, Rammer2018}, one obtains 
\begin{widetext}
\begin{subequations}
\begin{align}
&
\Pi^{\rm C}(\bm{q},z,z')
=
i\sum_{\bm{k}} \int_C dz_1 \int_C dz_2\,
G^{\rm C}_\up(\bm{k}^+,z,z_1)	
G^{\rm C}_\down(-\bm{k}^-,z,z_2)\,
\Lambda(\bm{q},z_1,z_2,z')
,\\
&
\Lambda(\bm{q},z,z',z'')
=
1 +\m{W}_{\rm imp}\sum_{\bm{k}}\int_C dz_1 \int_C dz_2\,
G^{\rm C}_\up(\bm{k}^+, z, z_1)
G^{\rm C}_\down(-\bm{k}^-, z', z_2)\,
\Lambda(\bm{q},z_1,z_2,z'').
\end{align}
\label{eq.app.contour.Pi}
\end{subequations}
Here, $\bm{k}^\pm = \bm{k} \pm \bm{q}/2$ and $\m{W}_{\rm imp} = n_{\rm imp} v_{\rm imp}^2$. $G^{\rm C}_\sigma$ is the dressed contour-ordered single-particle Green’s function, including the impurity self-energy as well as the self-energy arising from coupling to the normal-metal leads. The function $\Lambda$ is the three-point vertex correction arising from impurity scattering.

We now map the contour-time expressions in Eq.~\eqref{eq.app.contour.Pi} onto real-time functions. For clarity, we focus on the $n$th-order contribution $\Pi^{\rm C}_{n}$ with respect to the impurity-scattering strength $\m{W}_{\rm imp}$, which is depicted in Fig.~\ref{Fig.diagram}(a). The contour-ordered expression for $\Pi^{\rm C}_{n}$ has the form
\begin{align}
&
\Pi^{\rm C}_{n}(\bm{q}, z,z')
=
i\, \m{W}_{\rm imp}^n \sum_{\bm{k}_0,\cdots,\bm{k}_n}
\big[G_\up(\bm{k}_0^+) \circ G_\up(\bm{k}_1^+) \circ \cdots \circ G_\up(\bm{k}_n^+) \big](z,z')\,
\big[G_\down(-\bm{k}_0^-) \circ G_\down(-\bm{k}_1^-) \circ \cdots \circ G_\down(-\bm{k}_n^-) \big](z,z'),
\label{eq.app.PiC.n}
\end{align}
where we have introduced the convolution 
\begin{equation}
\big[G_\sigma(\bm{k}) \circ G_\sigma(\bm{k}')\big](z,z')
\equiv
\int_C dz_1\, G_\sigma(\bm{k}, z,z_1)\, 
G_\sigma(\bm{k}',z_1, z').
\end{equation}
The retarded component $\Pi^{\rm R}_n(\bm{q},t,t')$ of the $n$th-order contribution can be obtained by applying the Langreth rules~\cite{Stefanucci2013, Rammer2007, Haug2008} to Eq.~\eqref{eq.app.PiC.n}. Fourier transforming $\Pi^{\rm R}_n(\bm{q},t,t')$ with respect to the relative time $t-t'$, we obtain
\begin{align}
&
\Pi^{\rm R}_{n}(\bm{q},\nu)
=
\frac{i\, \m{W}_{\rm imp}^n}{2}\sum_{\bm{k}_0,\cdots,\bm{k}_n}
\int_{-\infty}^\infty \frac{d\omega}{2\pi}\,
\Big[
P_\up^{\rm R}(\bm{k}_0^+, \cdots, \bm{k}_n^+, \omega+\nu)\,
P_\down^{\rm K}(-\bm{k}_0^-, \cdots, -\bm{k}_n^-, -\omega)
\notag\\
&\hspace{0.5cm}+
P_\up^{\rm K}(\bm{k}_0^+, \cdots, \bm{k}_n^+, \omega+\nu)\,
P_\down^{\rm R}(-\bm{k}_0^-, \cdots, -\bm{k}_n^-, -\omega)
+
P_\up^{\rm R}(\bm{k}_0^+, \cdots, \bm{k}_n^+, \omega+\nu)\,
P_\down^{\rm R}(-\bm{k}_0^-, \cdots, -\bm{k}_n^-, -\omega)
\Big],
\label{eq.app.PiC.n.2}
\end{align}
where
\begin{equation}
P^{\rm X=R, A, K}_\sigma (\bm{k}_0, \cdots, \bm{k}_n, \omega)=
\big[G_\sigma(\bm{k}_0) G_\sigma(\bm{k}_1)\cdots G_\sigma(\bm{k}_n)\big]^{\rm X}(\omega).
\label{eq.def.Px}	
\end{equation}
Equation~\eqref{eq.def.Px} can be evaluated according to the standard multiplication rules of nonequilibrium Green’s functions~\cite{Stefanucci2013, Rammer2007, Haug2008}, giving 
\begin{subequations}
\begin{align}
P^{\rm R/A}_\sigma (\bm{k}_0, \cdots, \bm{k}_n, \omega)
&=
G^{\rm R/A}_\sigma(\bm{k}_0, \omega) 
G^{\rm R/A}_\sigma(\bm{k}_1, \omega) \cdots 
G^{\rm R/A}_\sigma(\bm{k}_n, \omega)
,\\[6pt]
P^{\rm K}_\sigma (\bm{k}_0, \cdots, \bm{k}_n, \omega)
&=
G^{\rm K}_\sigma(\bm{k}_0, \omega) 
G^{\rm A}_\sigma(\bm{k}_1, \omega) \cdots 
G^{\rm A}_\sigma(\bm{k}_n, \omega)
\notag\\
&\hspace{1.5cm} 
+
G^{\rm R}_\sigma(\bm{k}_0, \omega) 
G^{\rm K}_\sigma(\bm{k}_1, \omega) \cdots 
G^{\rm A}_\sigma(\bm{k}_n, \omega)
\notag\\
&\hspace{5cm}\vdots
\notag\\
&\hspace{1.5cm}
+
G^{\rm R}_\sigma(\bm{k}_0, \omega) 
G^{\rm R}_\sigma(\bm{k}_1, \omega) \cdots 
G^{\rm K}_\sigma(\bm{k}_n, \omega).
\label{eq.app.K.Langreth}
\end{align}
\label{eq.app.def.prod.G}
\end{subequations}
In the Keldysh component in Eq.~\eqref{eq.app.K.Langreth}, each term contains exactly one Keldysh Green’s function $G^{\rm K}_\sigma$, with all Green’s functions to its left taken as retarded components and all those to its right taken as advanced components.

Although the expansion of $P^{\rm X}_\sigma$ in Eq.~\eqref{eq.app.PiC.n.2} using Eq.~\eqref{eq.app.def.prod.G} appears to generate many terms, any term containing a factor of the form
\begin{equation}
\sum_{\bm{k}_j}
G^{\rm R}_\up(\bm{k}_j^+,\omega+\nu)
G^{\rm R}_\down(-\bm{k}_j^-,-\omega)
\end{equation}
actually vanishes after the $\bm{k}_j$ summation is carried out. As a result, only the following two terms survive:
\begin{align}
\Pi^{\rm R}_{n}(\bm{q},\nu)
&=
\frac{i\, \m{W}_{\rm imp}^n}{2} \sum_{\bm{k}_0,\cdots,\bm{k}_n}
\int_{-\infty}^\infty \frac{d\omega}{2\pi}
\notag\\
&\times
\bigg[
G^{\rm R}_\up(\bm{k}^+_0, \omega+\nu)\,
\big[
G_\up(\bm{k}^+_1)
\cdots
G_\up(\bm{k}^+_n)
\big]^{\rm R}(\omega+\nu)\,
G^{\rm K}_\down(-\bm{k}^-_0, -\omega)
\big[
G_\down(-\bm{k}^-_1)
\cdots
G_\down(-\bm{k}^-_n)
\big]^{\rm A}(-\omega)
\notag\\
&\hspace{0.5cm}+
G^{\rm K}_\up(\bm{k}^+_0, \omega+\nu)\,
\big[
G_\up(\bm{k}^+_1)
\cdots
G_\up(\bm{k}^+_n)
\big]^{\rm A}(\omega+\nu)\,
G^{\rm R}_\down(-\bm{k}^-_0, -\omega)
\big[
G_\down(-\bm{k}^-_1)
\cdots
G_\down(-\bm{k}^-_n)
\big]^{\rm R}(-\omega)
\bigg].
\label{eq.app.PiR.n.final}
\end{align}
We now sum up all orders in the impurity scattering strength $\m{W}_{\rm imp}$. Using Eq.~\eqref{eq.app.PiR.n.final}, we obtain
\begin{align}
\Pi^{\rm R}(\bm{q},\nu)
&=
\sum_{n=0}^\infty\, \Pi^{\rm R}_{n}(\bm{q},\nu)
\notag\\
&=
\frac{i}{2}
\sum_{\bm{k}}\int_{-\infty}^\infty \frac{d\omega}{2\pi}\,
\Bigg[
G^{\rm R}_\up(\bm{k}^+, \omega+\nu)
G^{\rm K}_\down(-\bm{k}^-, \omega+\nu)
\bigg[
1 + 
\m{W}_{\rm imp} \sum_{\bm{k}_1}
G^{\rm R}_\up(\bm{k}_1^+, \omega+\nu)
G^{\rm A}_\down(-\bm{k}_1^-, -\omega)
+
\cdots
\bigg]
\notag\\[4pt]
&\hspace{3.5cm}+
G^{\rm K}_\up(\bm{k}^+, \omega+\nu)
G^{\rm R}_\down(-\bm{k}^-, \omega+\nu)
\bigg[
1 + 
\m{W}_{\rm imp} \sum_{\bm{k}_1}
G^{\rm A}_\up(\bm{k}_1^+, \omega+\nu)
G^{\rm R}_\down(-\bm{k}_1^-, -\omega)
+
\cdots
\bigg]
\Bigg]
\notag\\
&=
\frac{i}{2}
\sum_{\bm{k}}\int_{-\infty}^\infty \frac{d\omega}{2\pi}\,
\Big[
G^{\rm R}_\up(\bm{k}^+, \omega+\nu)
G^{\rm K}_\down(-\bm{k}^-, \omega+\nu)\,
\Lambda(\bm{q}, \omega+\nu, -\omega)
\notag\\
&\hspace{4cm}+
G^{\rm K}_\up(\bm{k}^+, \omega+\nu)
G^{\rm R}_\down(-\bm{k}^-, \omega+\nu)\,
\Lambda^*(\bm{q}, \omega+\nu, -\omega)
\Big],
\end{align}
which gives Eqs.~\eqref{eq.PiR} and \eqref{eq.BSeq}.
\end{widetext}

\section{Derivation of Eq.~\eqref{eq.rho.CI.noneq}
\label{sec.app.derivation.CI}}

We derive Eq.~\eqref{eq.rho.CI.noneq}. In this Appendix, we focus on the clean-limit case without impurity scattering. Since the spin structure becomes trivial in the absence of magnetic impurity scattering, it is sufficient to work with the two-component Nambu field,
\begin{equation}
\bm{\Psi}^\dagger_{\bm{k}} =
\begin{pmatrix}
a^\dagger_{\bm{k},\up} &
a_{-\bm{k},\down}
\end{pmatrix},
\label{eq.Namnbu.2}
\end{equation}
instead of the four-component one in Eq.~\eqref{eq.Namnbu.4}. In the following, the retarded, advanced, and Keldysh Green's functions in Eq.~\eqref{eq.def.Nambu.Green} are defined by using Eq.~\eqref{eq.Namnbu.2}. For later convenience, we also introduce the lesser Green's function,
\begin{equation}
\bm{G}^<(\bm{k},\omega)=
-i \int_{-\infty}^\infty dt\, e^{i\omega t}
\braket{\bm{\Psi}^\dagger_{\bm{k}}(0)\, \bm{\Psi}_{\bm{k}}(t)},
\end{equation}
which is related to $\bm{G}^{\rm R,A,K}$ as
\begin{equation}
\bm{G}^<(\bm{k},\omega) =
\frac{1}{2}\Big[
-\bm{G}^{\rm R}(\bm{k},\omega)
+\bm{G}^{\rm A}(\bm{k},\omega)
+\bm{G}^{\rm K}(\bm{k},\omega)
\Big].
\label{eq.G<.RAK}
\end{equation}
Using the $(1,1)$ component $G^<_{11}$ of the lesser Green's function, the total charge density $\rho$ $(=\rho_\up+\rho_\down=2\rho_\up)$ in the superconductor is given by
\begin{equation}
\rho
=
-2ie \sum_{\bm{k}}
\int_{-\infty}^\infty \frac{d\omega}{2\pi}\,
G^<_{11}(\bm{k}, \omega).
\label{eq.G<11.rho}
\end{equation}

To evaluate the lesser Green's function in order to calculate $\rho$ in Eq.~\eqref{eq.G<11.rho}, we first consider the retarded Green's function in Eq.~\eqref{eq.dressed.GR}. In the clean limit ($\bm{\Sigma}^{{\rm X}={\rm R,A,K}}_{\rm imp}=\bm{\Sigma}^{\rm X}_{\rm mag}=0$), one finds
\begin{align}
\bm{G}^{\rm R}(\bm{k},\omega)
&=
\frac{\omega_+\,\tau_0+
[\xi_{\bm{k}}+e\varphi]\, \tau_3+
\Delta\,\tau_1}
{\omega_+^2 -[\xi_{\bm{k}}+e\varphi]^2 -\Delta^2}
\notag\\
&=
\frac{1}{\omega + i\gamma/2 - E_{\bm{k}}}
\bm{P}_{\bm{k}}^+
+
\frac{1}{\omega + i\gamma/2 + E_{\bm{k}}}
\bm{P}_{\bm{k}}^-,
\label{eq.GR.clean.2}
\end{align}
where
\begin{equation}
\bm{P}^+_{\bm{k}}=
\tau_0 -\bm{P}^-_{\bm{k}} =
\begin{pmatrix}
u^2_{\bm{k}} & -u_{\bm{k}} v_{\bm{k}} \\[4pt]
- u_{\bm{k}} v_{\bm{k}} & v^2_{\bm{k}}
\end{pmatrix}.
\end{equation}
Here, the Bogoliubov excitation energy $E_{\bm{k}}$ and the coherence factors, $u_{\bm{k}}$ and $v_{\bm{k}}$, are given in Eqs.~\eqref{eq.Ek.phi} and \eqref{eq.ukvk.phi}, respectively.

For the Keldysh component, the self-energy in Eq.~\eqref{eq.self.lead.K.Nambu} is given by, in the two-component Nambu representation,
\begin{align}
&
\bm{\Sigma}^{\rm K}_{\rm lead}(\bm{k},\omega)
=
-i \sum_{\alpha=1,2} \gamma_\alpha
\notag\\
&
\scalebox{0.9}{$\displaystyle
\times
\begin{pmatrix}
1-2f(\omega +\mu_{\rm pair} -eV\delta_{\alpha,1})
& 0 \\
0 &
1-2f(\omega -\mu_{\rm pair} +eV\delta_{\alpha,1})
\end{pmatrix}$}.
\label{eq.SigmaK.lead.2}	
\end{align}
Using Eqs.~\eqref{eq.GR.clean.2}, \eqref{eq.SigmaK.lead.2}, and
$\bm{\Sigma}^{\rm K}_{\rm int}=0$, we can obtain the Keldysh Green's function as
\begin{align}
\bm{G}^{\rm K}(\bm{k},\omega)
&=
\bm{G}^{\rm R}(\bm{k},\omega)\,
\bm{\Sigma}^{\rm K}_{\rm lead}(\bm{k},\omega)\,
\bm{G}^{\rm A}(\bm{k},\omega)
\notag\\[4pt]
&=
i\gamma
\bigg[
\frac{C^-_{\bm{k}}(\omega)}{D^{--}_{\bm{k}}(\omega)}\,
\bm{P}^{+}_{\bm{k}}
+
\frac{C^+_{\bm{k}}(\omega)}{D^{++}_{\bm{k}}(\omega)}\,
\bm{P}^{-}_{\bm{k}}
\notag\\
&\hspace{0.7cm}
+
\frac{S_{\bm{k}}(\omega)}{D^{-+}_{\bm{k}}(\omega)}\,
\bm{M}^{+}_{\bm{k}}
+
\frac{S_{\bm{k}}(\omega)}{D^{+-}_{\bm{k}}(\omega)}\,
\bm{M}^{-}_{\bm{k}}
\bigg],
\label{eq.app.GK.asy}
\end{align}
where
\begin{subequations}
\begin{align}
&
D^{\eta \eta'}_{\bm{k}}(\omega)
=
\big[\omega +i\gamma/2 +\eta E_{\bm{k}}\big]
\big[\omega -i\gamma/2 +\eta' E_{\bm{k}}\big]
,\\[4pt]
&
C^\eta_{\bm{k}}(\omega)
=
2 u^2_{\bm{k}}\, f^\eta_{\rm neq}(\omega) +2 v^2_{\bm{k}}\, f^{-\eta}_{\rm neq}(\omega) -1
\label{eq.def.Ceta}
,\\[4pt]
&
S_{\bm{k}}(\omega)
=
2 u_{\bm{k}} v_{\bm{k}} 
\big[f^-_{\rm neq}(\omega) -f^+_{\rm neq}(\omega)\big]
\label{eq.def.Sk}
,\\[4pt]
&
\bm{M}^+_{\bm{k}}= i\tau_2 +\bm{M}^-_{\bm{k}}=
\begin{pmatrix}
u_{\bm{k}} v_{\bm{k}} & u^2_{\bm{k}} \\[4pt]
-v^2_{\bm{k}} & -u_{\bm{k}} v_{\bm{k}} 
\end{pmatrix},
\end{align}
\end{subequations}
with $f^\pm_{\rm neq}(\omega)$ being defined in Eq.~\eqref{eq.def.fneq.pm}.

The lesser Green's function $\bm{G}^<$ is thus obtained from
Eqs.~\eqref{eq.G<.RAK}, \eqref{eq.GR.clean.2}, and \eqref{eq.app.GK.asy}.
Substituting the $(1,1)$ component $G^<_{11}$ into
Eq.~\eqref{eq.G<11.rho}, we obtain
\begin{align}
\rho
&=
2e \sum_{\bm{k}}
\int_{-\infty}^\infty \frac{d\omega}{2\pi}
\Bigg[
u^2_{\bm{k}}
\big[
u^2_{\bm{k}}\, f^-_{\rm neq}(\omega) +
v^2_{\bm{k}}\, f^+_{\rm neq}(\omega)
\big]
\frac{\gamma}{D^{--}_{\bm{k}}(\omega)}
\notag\\[4pt]
&\hspace{0.5cm}
+
v^2_{\bm{k}}
\big[
u^2_{\bm{k}}\, f^+_{\rm neq}(\omega) +
v^2_{\bm{k}}\, f^-_{\rm neq}(\omega)
\big]
\frac{\gamma}{D^{++}_{\bm{k}}(\omega)}
\notag\\
&\hspace{0.5cm}
+
u^2_{\bm{k}} v^2_{\bm{k}}
\big[f^-_{\rm neq}(\omega)-f^+_{\rm neq}(\omega)\big]
\left[
\frac{\gamma}{D^{-+}_{\bm{k}}(\omega)} +
\frac{\gamma}{D^{+-}_{\bm{k}}(\omega)}
\right]
\Bigg].
\label{eq.rho.complicated}
\end{align}

To obtain a physically transparent expression, we take the limit
$\gamma\to +0$ in Eq.~\eqref{eq.rho.complicated} while keeping the ratio $\gamma_1/\gamma_2$ fixed.
In this limit, by using
\begin{align}
&
\lim_{\gamma\to +0}
\frac{1}{2\pi}
\frac{\gamma}{D^{\eta\eta}_{\bm{k}}(\omega)}
=
\delta\!\left(\omega+\eta E_{\bm{k}}\right),
\label{eq.delta.simple}
\\[4pt]
&
\lim_{\gamma\to +0}
\frac{1}{2\pi}
\left[
\frac{\gamma}{D^{-+}_{\bm{k}}(\omega)}
+
\frac{\gamma}{D^{+-}_{\bm{k}}(\omega)}
\right]
\notag\\[4pt]
&\hspace{0.8cm}
=
\left[1-\frac{E_{\bm{k}}}{\omega}\right]
\delta\!\left(\omega- E_{\bm{k}}\right)
+
\left[1+\frac{E_{\bm{k}}}{\omega}\right]
\delta\!\left(\omega+ E_{\bm{k}}\right),
\label{eq.delta.mixed}
\end{align}
together with the identity
\begin{equation}
f^\pm_{\rm neq}(-E_{\bm{k}})
=
1-f^\mp_{\rm neq}(E_{\bm{k}}),
\label{eq.f.identity}
\end{equation}
Eq.~\eqref{eq.rho.complicated} can be rearranged as
\begin{align}
\rho
&=
2e\sum_{\bm{k}}
\Big[
u^2_{\bm{k}}\big[
u^2_{\bm{k}} f^-_{\rm neq}(E_{\bm{k}})
+
v^2_{\bm{k}} f^+_{\rm neq}(E_{\bm{k}})
\big]
\notag\\
&\hspace{2.3cm}
+
v^2_{\bm{k}}\big[
u^2_{\bm{k}} f^+_{\rm neq}(-E_{\bm{k}})
+
v^2_{\bm{k}} f^-_{\rm neq}(-E_{\bm{k}})
\big]
\Big]
\notag\\[4pt]
&=
2e\sum_{\bm k}\Big[
\big[u_{\bm k}^4-v_{\bm k}^4\big] f^{\rm s}_{\rm neq}(E_{\bm k})
\notag\\
&\hspace{2.3cm}
-
\big[u_{\bm k}^2-v_{\bm k}^2\big]^2 f^{\rm a}_{\rm neq}(E_{\bm k})
+
v_{\bm k}^2
\Big]
\notag\\[4pt]
&=
2e\sum_{\bm k}
\Big[
\big[u_{\bm k}^2-v_{\bm k}^2\big] f^{\rm s}_{\rm neq}(E_{\bm k})
-
\big[u_{\bm k}^2-v_{\bm k}^2\big]^2 f^{\rm a}_{\rm neq}(E_{\bm k})
\Big]
\notag\\
&\hspace{2.3cm}
+
2e\sum_{\bm{k}} v^2_{\bm{k}}
\notag\\[4pt]
&=
\underbrace{
2\sum_{\bm{k}} q_{\bm{k}}
\Big[
f^{\rm s}_{\rm neq}(E_{\bm{k}})
-
\frac{\xi_{\bm{k}} +e\varphi}{E_{\bm{k}}}
f^{\rm a}_{\rm neq}(E_{\bm{k}})
\Big]
}_{\equiv\,\rho_{\rm qp}}
+
\underbrace{
2e\sum_{\bm{k}} v^2_{\bm{k}}
}_{\equiv\,\rho_{\rm pair}},
\label{eq.app.rho.neq}
\end{align}
where $f^{\rm s/a}_{\rm neq}(E_{\bm{k}})$ is defined in Eq.~\eqref{eq.def.fas}. Equation~\eqref{eq.app.rho.neq} reproduces Eq.~\eqref{eq.rho.CI.noneq}.

\section{Proof of Eq.~\eqref{eq.GK_full_h} \label{sec.app.proof}}

We prove Eq.~\eqref{eq.GK_full_h}, which shows that the nonequilibrium distribution in the Keldysh Green's function remains unchanged by nonmagnetic impurity scattering. We start from the general ansatz for the Keldysh component of the impurity-dressed Green's function~\cite{Rammer2007, Kopnin2001},
\begin{equation}
\bm{G}^{\rm K}(\bm{k},\omega)
=
\bm{G}^{\rm R}(\bm{k},\omega)\, \bm{h}_{\rm imp}(\omega)
-
\bm{h}_{\rm imp}(\omega)\, \bm{G}^{\rm A}(\bm{k},\omega).
\label{eq.GKimp.ansatz}
\end{equation}
Our goal is to show that $\bm{h}_{\rm imp}(\omega)$ coincides with the distribution function $\bm{h}(\omega)$ appearing in the clean-limit Keldysh Green's function $\bm{G}^{\rm K}_{\rm clean}$ in Eq.~\eqref{eq.GK_clean_h}.

To this end, we consider the Keldysh sector of the Dyson equation~\eqref{eq.Dyson_imp}, having the form~\cite{Stefanucci2013, Rammer2007, Haug2008}
\begin{equation}
\bm{G}^{\rm K}
=
\big[\bm{1} +\bm{G}^{\rm R} \bm{\Sigma}^{\rm R}_{\rm imp} \big]\,
\bm{G}^{\rm K}_{\rm clean}\,
\big[\bm{1} +\bm{\Sigma}^{\rm A}_{\rm imp} \bm{G}^{\rm A} \big]
+
\bm{G}^{\rm R}\, \bm{\Sigma}^{\rm K}_{\rm imp}\,
\bm{G}^{\rm A}.
\label{eq.app.Dyson12}
\end{equation}
Hereafter, we suppress the arguments $(\bm{k},\omega)$ for notational simplicity. From Eq.~\eqref{eq.self.imp.Nambu} together with the ansatz in Eq.~\eqref{eq.GKimp.ansatz}, the impurity self-energy $\bm{\Sigma}^{\rm K}_{\rm imp}$ is evaluated as
\begin{align}
&
\bm{\Sigma}^{\rm K}_{\rm imp}
\notag\\
&
=
n_{\rm imp} u_{\rm imp}^2
\sum_{\bm{k}}
\big(\tau_3 \otimes \sigma_0\big)\,
\big[
\bm{G}^{\rm R}\, \bm{h}_{\rm imp}
-
\bm{h}_{\rm imp}\, \bm{G}^{\rm A}
\big]\,
\big(\tau_3 \otimes \sigma_0\big)
\notag\\
&=
\bm{\Sigma}^{\rm R}_{\rm imp}\, \bm{h}_{\rm imp}
-
\bm{h}_{\rm imp}\, \bm{\Sigma}^{\rm A}_{\rm imp}.
\label{eq.app.sigK.imp}
\end{align}
Substituting Eqs.~\eqref{eq.GK_clean_h}, \eqref{eq.GKimp.ansatz}, and \eqref{eq.app.sigK.imp} into the Keldysh equation~\eqref{eq.app.Dyson12}, we obtain
\begin{align}
&
\bm{G}^{\rm R}\, \bm{h}_{\rm imp}
-
\bm{h}_{\rm imp}\, \bm{G}^{\rm A}
\notag\\[4pt]
&=
\big[\bm{1} +\bm{G}^{\rm R} \bm{\Sigma}^{\rm R}_{\rm imp} \big]
\big[
\bm{G}^{\rm R}_{\rm clean} \bm{h}
-
\bm{h}\, \bm{G}^{\rm A}_{\rm clean}
\big]
\big[\bm{1} +\bm{\Sigma}^{\rm A}_{\rm imp} \bm{G}^{\rm A} \big]
\notag\\
&\quad
+
\bm{G}^{\rm R}
\big[
\bm{\Sigma}^{\rm R}_{\rm imp}\, \bm{h}_{\rm imp}
-
\bm{h}_{\rm imp}\, \bm{\Sigma}^{\rm A}_{\rm imp}
\big]
\bm{G}^{\rm A}.
\label{eq.app.proof1}
\end{align}
By using the Dyson equations for the retarded and advanced Green's functions,
\begin{align}
\bm{G}^{\rm R} 
&=
\bm{G}^{\rm R}_{\rm clean}
+
\bm{G}^{\rm R}
\bm{\Sigma}^{\rm R}_{\rm imp}
\bm{G}^{\rm R}_{\rm clean},
\\
\bm{G}^{\rm A} 
&=
\bm{G}^{\rm A}_{\rm clean}
+
\bm{G}^{\rm A}_{\rm clean}
\bm{\Sigma}^{\rm A}_{\rm imp}
\bm{G}^{\rm A},
\end{align}
Eq.~\eqref{eq.app.proof1} can be rewritten as
\begin{align}
&
\bm{G}^{\rm R}\, \bm{h}_{\rm imp}
-
\bm{h}_{\rm imp}\, \bm{G}^{\rm A}
=
\bm{G}^{\rm R}\, \bm{h}
-
\bm{h}\, \bm{G}^{\rm A}
\notag\\[4pt]
&\quad
+
\bm{G}^{\rm R}
\bm{\Sigma}^{\rm R}_{\rm imp}
\big[\bm{h}_{\rm imp}-\bm{h}\big]
\bm{G}^{\rm A}
-
\bm{G}^{\rm R}
\big[\bm{h}_{\rm imp}-\bm{h}\big]
\bm{\Sigma}^{\rm A}_{\rm imp}
\bm{G}^{\rm A}.
\label{eq.app.proof}
\end{align}
Equation~\eqref{eq.app.proof} immediately concludes that $\bm{h}_{\rm imp}(\omega) = \bm{h}(\omega)$ is required for consistency.

\bibliography{NSN}
\end{document}